\documentclass[preprint,12pt]{aastex}  

\def\cm{\ifmmode {\rm cm}^{-1} \else cm$^{-1}$ \fi}
\def\s{\ifmmode {\rm s}^{-1} \else s$^{-1}$ \fi}
\def\cc{\ifmmode {\rm cm}^{-3} \else cm$^{-3}$ \fi}
\def\cs{\ifmmode {\rm cm}^{-2} \else cm$^{-2}$ \fi}
\def\g{\ifmmode \gamma \else $\gamma$\fi}
\def\G{\ifmmode \Gamma \else $\Gamma$\fi}
\def\Gs{\ifmmode \Gamma~ \else $\Gamma~$\fi}
\def\Ka{K$\alpha$~}

\def\gc{\ifmmode \gamma_{\rm c} \else $\gamma_{\rm c}$ \fi}
\def\sw{Schwarzschild~}
\def\gsim{\mathrel{\raise.5ex\hbox{$>$}\mkern-14mu
             \lower0.6ex\hbox{$\sim$}}}

\def\lsim{\mathrel{\raise.4ex\hbox{$<$}\mkern-14mu
             \lower0.6ex\hbox{$\sim$}}}

\def\simless{\mathbin{\lower 3pt\hbox
     {$\rlap{\raise 5pt\hbox{$\char'074$}}\mathchar"7218$}}}   
\def\simmore{\mathbin{\lower 3pt\hbox
     {$\rlap{\raise 5pt\hbox{$\char'076$}}\mathchar"7218$}}}   
\def\4u{4U 1728--34}

\received{} \accepted{}
\slugcomment{Accepted to ApJ, April 16, 2007}

\begin{document}

\title{Accretion Disk Illumination in Schwarzschild and Kerr Geometries:
Fitting Formulae }


\author{ }
%
%
\author{Keigo Fukumura \& Demosthenes Kazanas}
\affil{Astrophysics Science Division, NASA Goddard Space Flight
Center, Code 663, Greenbelt, MD 20771}
\email{fukumura@milkyway.gsfc.nasa.gov, Demos.Kazanas-1@nasa.gov}
%
%
%
%
%

\begin{abstract}

\baselineskip=15pt

We describe the methodology and compute the illumination of
geometrically thin accretion disks around black holes of arbitrary
spin parameter $a$ exposed to the radiation of a point-like,
isotropic source at arbitrary height above the disk on its symmetry
axis. We then provide analytic fitting formulae for the illumination
as a function of the source height $h$ and the black hole angular
momentum $a$. We find that for a source on the disk symmetry axis
and $h/M > 3$, the main effect of the parameter $a$ is allowing the
disk to extend to smaller radii (approaching $r/M \rightarrow 1$ as
$a/M \rightarrow 1$) and thus allow the illumination of regions of
much higher rotational velocity and redshift. We also compute the
illumination profiles for anisotropic emission associated with the
motion of the source relative to the accretion disk and present the
fractions of photons absorbed by the black hole, intercepted by the
disk or escaping to infinity for both isotropic and anisotropic
emission for $a/M=0$ and $a/M=0.99$. As the anisotropy (of a source
approaching the disk) increases the illumination profile reduces
(approximately) to a single power-law, whose index, $q$, because of
absorption of the beamed photons by the black hole, saturates to a
value no higher than $q \gtrsim 3$. Finally, we compute the
fluorescence Fe line profiles associated with the specific
illumination and compare them among various cases.

\end{abstract}


\baselineskip=15pt

\section{Introduction}

The ``standard" geometry of accreting active galactic nuclei (AGNs)
(and for that matter also galactic black hole candidates) consists
of an optically thick, geometrically thin accretion disk that
extends to the radius of the innermost stable circular orbit (ISCO)
surrounded by a hot ($T \sim 10^8-10^9$ K), X-ray emitting corona.
Coronal X-rays are reprocessed on the surface of the much cooler
disk ($T \sim 10^4-10^6$ K) to produce fluorescent Fe \Ka emission
line at 6.4 keV as well as a reflection component; the latter
constitutes a broad peak near $E \simeq 30$ keV  resulting from the
reprocessing of the power law coronal emission on the colder disk by
reducing its flux by photoelectric absorption at low energies and by
down-Comptonization at high energies.

The intrinsically narrow shape and the well known energy (6.4 keV)
of the \Ka line make it a unique diagnostic of the inner regions of
accretion flows around black holes: An accretion disk such as those
described above is expected to produce line profiles of unique
shapes because of the large Doppler motion and redshifts associated
with emission from the innermost radii of such disks \citep[see
however][for an alternative view]{Titar03}. Model fluorescence \Ka
profiles have been computed for disks around either \sw
\citep{Fab89} or Kerr
\citep[e.g.,][]{Laor91,Beckwith04,Dovciak04,Reynolds06} black holes,
while more general treatment including the reflection component has
also been discussed \citep[e.g.,][]{George91,Matt91}.

The advent of observation has subsequently vindicated these
considerations with the discovery of broad iron \Ka lines
\citep[e.g.,][]{Tanaka95,Nandra97}. Of particular interest has been
the \Ka line associated with the AGN MCG --6-30-15, which, because
of its very broad red wing, indicated emission by matter interior to
the radius of the ISCO of a \sw black hole ($r/M = 6$ where $M$ is
the black hole mass), thereby implying emission by an accretion disk
around a highly rotating Kerr black hole \citep[][]{Iwasawa96}.
Observations of the same object by the much larger area of {\it
XMM-Newton} increased the significance of these observations and
confirmed the presence of a very broad line \citep{Wilms01,Fab02},
however the low sensitivity of the instrument above $\gsim 10$ keV
has made difficult the determination of the underlying continuum and
therefore the true width of the line. More recently, observations of
MCG --6-30-15 by \citet{Miniutti07} with {\it Suzaku}, with improved
sensitivity at higher energies over {\it XMM-Newton}, were able to
confirm that the Fe \Ka line in this object is indeed as broad as it
was originally claimed.

Clearly, the profile of the Fe \Ka emission depends, besides the
geometric properties of the accretion disk, also on its kinematic
properties and on its illumination by the X-ray continuum source.
While the aforementioned treatments, as well as most of similar
treatments in the literature assume axisymmetry and Keplerian motion
for the plasma in the disk, these assumptions are not guaranteed by
any fundamental principle. Non-axisymmetric structures and/or
non-negligible radial velocity in the disk can yield lines of
different shapes as shown by some authors
\citep[e.g.,][]{Ruszkowski00,Yu00,FT04}. Finally, even in
axisymmetric geometries, different illumination laws can yield line
profiles of different shapes, as originally shown by \citet{Laor91}.
For computational simplicity, a (broken) power-law form ($r^{-q}$
where $q>0$) for the illumination function is generally assumed
\citep[e.g.,][]{Fab02}. \citet{Laor91} has presented line profiles
both with such illumination laws as well as with that of an infinite
plane by a point source  at height $h$ in flat space, i.e., $F(x) =
h/[4 \pi(h^2 +x^2)^{3/2}]$, where $x$ is the distance on the disk
plane from the foot of the vertical line from the source to the
plane.

The sensitivity of the line profiles on this parameter \citep[see][
Fig.~2]{Laor91} has led us to take a closer look into this issue.
This issue has been addressed to a certain extent by \citet{RB97}
who computed the illumination for a \sw black hole assuming an
isotropic source at a given height, with emphasis on the
contribution to the line emission by matter in free fall region
(i.e., plunging region), interior to the ISCO at $r/M = 6$. The
question we are set to study in this note is the illumination law of
an accretion disk by isotropic and non-isotropic point sources at
arbitrary values of the height $h$ and for arbitrary values of the
black hole spin parameter $a$. Our goal is to produce fitting
formulae for the most generic cases which can be easily incorporated
into existing numerical/software codes to produce more realistic
line profiles for given source configurations. We place some
emphasis on the issue of isotropy which changes in the presence of
strong gravitational field. We also provide examples of anisotropic
illumination by assuming that the source is moving instantaneously
and that all anisotropy is due to the relativistic beaming of its
photons.

In \S 2 we outline the setup of the problem and the method followed
for the case of a Schwarzschild black hole, while we defer
discussion of the Kerr geometry to the Appendix. In \S 3 we present
the results of our calculations along with their fitting formulae in
graphic form for comparison with the direct calculation and provide
a few examples of the resulting Fe \Ka line profiles. Finally in \S
4 we conclude by summarizing and evaluating our results.

\section{Formulation of the Problem}

\begin{figure}[t]
\epsscale{.50} \plotone{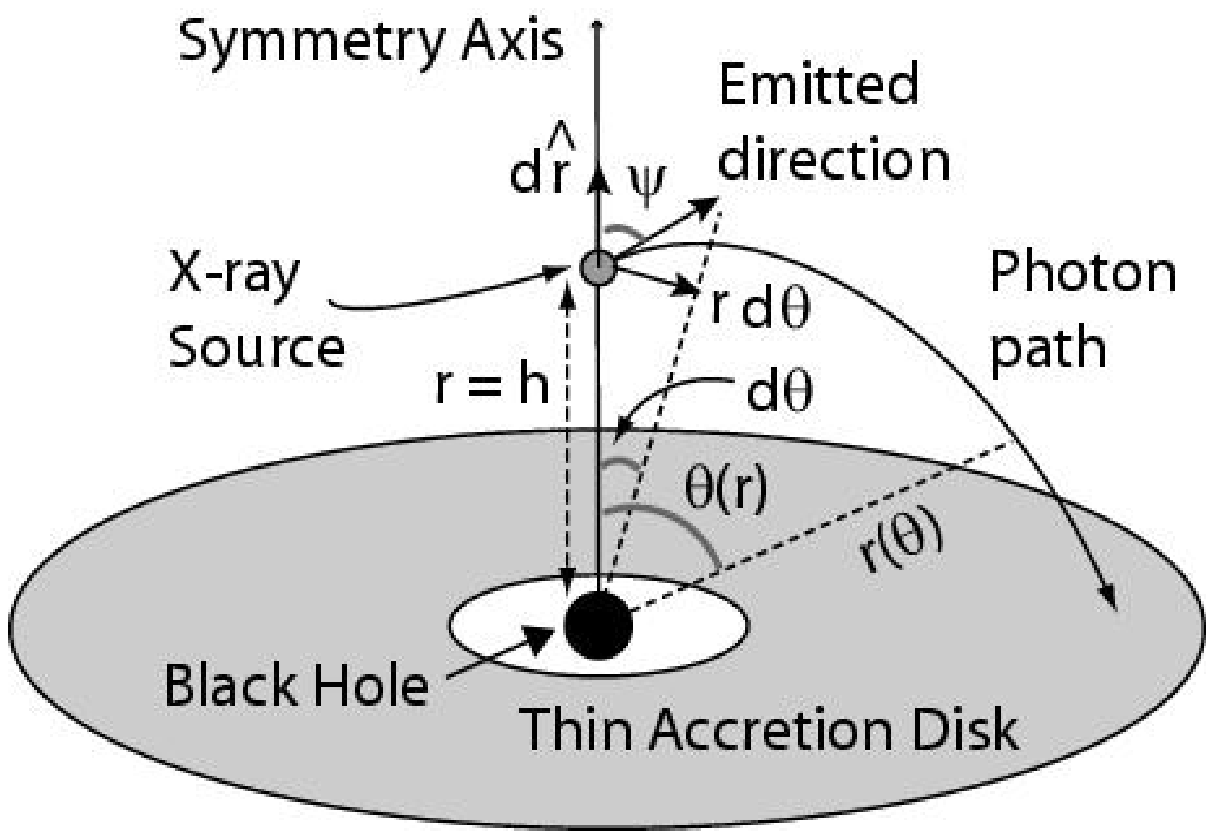} \caption{Schematic illustration of
our model. An accretion disk is illuminated by a point-like X-ray
source at height $h$ on the symmetry axis. \label{fig:geometry}}
\end{figure}

We consider an X-ray source located at a height $h$ above the plane
of an accretion disk and along its axis of symmetry as shown in
Figure~\ref{fig:geometry}. A black hole is located at the origin of
the coordinate system, with the accretion disk extending to the
ISCO. The latter depends on the spin parameter of the black hole,
ranging from $r/M=6$ for a \sw black hole ($a/M = 0$) to $r/M=1$ as
$a/M \rightarrow 1$. We also assume the X-ray emission to be
isotropic in the local frame of the X-ray source.


In this part of our treatment we follow the exposition given in
\citet{Chandrasekhar83} in formulating the problem of illumination
and the method followed; we restrict ourselves to discussing the \sw
case only, deferring the discussion of the Kerr geometry to the
Appendix. We adopt the usual geometrized units in the Kerr geometry
in which $G=c=1$ where $G$ and $c$ are the gravitational constant
and the speed of light, respectively. Thus, distance (length) is
scaled by the black hole mass $M$ throughout the paper. The event
horizon is then expressed as $r_{\rm{H}} \equiv M+\sqrt{M^2-a^2}
\rightarrow 2M$ as $a/M \rightarrow 1$. In the \sw case the radial
equation of the photon orbit is
\begin{equation}
\left(\frac{dr}{d\tau}\right)^2 = E^2 - \frac{L^2}{r^2}\left(1-
\frac{2M}{r}\right) \ , \label{eqmotion}
\end{equation}
to be considered along with the angular momentum $L$ and energy $E$
integrals of motion
\begin{equation}
r^2 \frac{d \theta}{d \tau} = L  ~~~~  {\rm and} ~~~~ \left( 1 -
\frac{2M}{r} \right) \frac{dt}{d \tau} = E \, . \label{integral}
\end{equation}
These, through the redefinition of the null affine parameter $\tau$
to $ d\tilde \tau \equiv E d \tau$, read
\begin{equation}
r^2 \frac{d \theta}{d \tilde \tau} = b  ~~~~  {\rm and} ~~~~ \left(
1 - \frac{2M}{r} \right) \frac{dt}{d \tilde \tau} = 1 \ ,
\label{integral2}
\end{equation}
where $b \equiv L/E$ is the impact parameter of the orbit, its
unique degree of freedom. In the remaining of the paper we drop the
tilde in the definition of $\tau$.

On dividing equation~(\ref{eqmotion}) by $r^4 (d \theta/d \tau)^2
=L^2$ one obtains the equation for the photon orbits
\begin{equation}
\left(\frac{1}{r^2}\frac{d r}{d \theta} \right)^2 = \frac{1}{b^2} -
\frac{1}{R^2(r)} \ , \label{eqorbit}
\end{equation}
where $R^2(r) \equiv r^2/(1- 2M/r)$.

As shown in Figure~\ref{fig:geometry} we consider photons emitted by
the source at (local) polar angles $\psi$ defined by the expression
\citep[see][p.127]{Chandrasekhar83}
\begin{equation}
{\rm cot} \psi = \frac{1}{r}\frac{d \tilde r}{d
\theta}=\frac{1}{r(1-2M/r)^{1/2}}\frac{d r}{d \theta} \ ,
\label{angle}
\end{equation}
where $d \tilde r=(1-2M/r)^{-1/2} dr$ is the element of proper
length in the radial direction and $r$ and $\theta$ are the usual
Boyer-Lindquist coordinates of the \sw geometry. We use the above
equation (\ref{angle}) for the definition of isotropic photon
emission at the local frame of the source, and so we choose their
distribution to be uniform in the angle $\psi$ (i.e., we set $\Delta
\psi = \rm{constant}$). Using the equation of the photon orbits
[equation (\ref{eqorbit})] and the definition of $R(r)$, the above
expression for the angle $\psi$ can be rewritten as
\begin{equation}
{\rm cot}^2 \psi =  \frac{R^2(r)}{b^2} - 1 ~~~~{\rm or} ~~~~ b^2 =\;
 R^2(r) \;{\rm sin^2} \psi \;= \; \frac{r^2}{(1-2M/r)}\;{\rm sin^2}
 \psi \ .
\label{angle2}
\end{equation}
%
The above equation~(\ref{angle2}) provides the necessary relation
between the photon impact parameter $b$ and its local emission angle
$\psi$.

The photon trajectories of a (locally) isotropic emission are
determined by choosing the value of $\psi$ uniformly between $0$ and
$\pi$ and then using equation~(\ref{angle2}) to determine the value
of the photon's impact parameter $b$ which, substituted in
equation~(\ref{eqorbit}), provides the associated photon orbits.
However, because the photon orbit equation involves the square of
the derivative $dr /d \theta$, we have found it more convenient to
use the second-order equation for $r$ supplemented with a value for
the derivative of the radial coordinate, $\dot r ~(\equiv
dr/d\tau)$, to be used as an initial condition. Differentiating
equation~(\ref{eqmotion}) with respect to $\tau$ and using
equation~(\ref{angle2}) we obtain the following expressions for the
equation of motion and the initial condition on $\dot r$
\begin{equation}
\ddot r = \frac{b^2}{r^3} \left(1 - \frac{3M}{r} \right) ~~~~ {\rm
and} ~~~~ \dot r ^2 = 1 - \frac{b^2}{R^2(r)} = 1 - \frac{b^2}{r^2}
\left(1 - \frac{2M}{r} \right) =  {\rm cos^2} \psi \ . \label{rdot}
\end{equation}
Cast in the above form, the photon equation of motion makes apparent
the significance of the $r/M=3$ surface for photon circular orbits
in the \sw geometry; the second relation then indicates that,
because of the single degree of freedom of the photon orbits, the
value of $\dot r$ necessary for their specification is uniquely
determined by the value of the impact parameter $b$ or the angle
$\psi$ through this relation.

Our procedure for the determination of the orbits is the following:
We consider a large number of orbits determined by the photon
emission angle $\psi$ chosen uniformly between $0$ and $\pi$. Using
equations (\ref{angle2}) and (\ref{rdot}) we obtain for each value
of $\psi$ the values for $b$ and $\dot r$ which are used to
integrate the second-order photon equation of motion along with that
for the angular momentum, $d \theta / d \tau = b / r^2$. The
integration proceeds from an initial value $\theta = 0$ (the
location of the source on the symmetry axis) until the angle
$\theta$ reaches the value $\theta = \pi/2$ (the position of the
equatorial plane where the accretion disk lies) at which point we
note the value of the radial coordinate $r$ (now on the surface of
the disk) at which the photon is located. Because we launch photons
only on the poloidal plane from a source located on the symmetry
axis of the source -- black hole configuration, in order to take into account
the photons emitted in the azimuthal direction, we weigh each photon
intercepting the accretion disk by sin$\psi$, the sine of the local
polar photon emission  angle.
We collect all the (weighted) photons arriving at the disk plane
(i.e., $\theta=\pi/2$) at a given radial interval, from  the horizon
$r_{\rm{H}}$ to $r/M= 2000$ which is then converted to photons per
unit area to produce the disk illumination.

\section{Results}

Following the procedure outlined above for a \sw black hole and for
a Kerr black hole in the Appendix we have produced the illumination
of accretion disks for a variety of values for the black hole spin
parameter $a$ and height of the source, $h$, expressed in units of
the black hole mass $M$. Our results along with the fitting formulae
appropriate for a most generic case, namely that of an isotropic
point source on the symmetry axis, are given in the next subsection
(\S 3.1). The effects of a locally anisotropic emission are discussed
in the following subsection (\S 3.2), while we conclude this section by
presenting the line profiles that result from the corresponding
model illumination laws (\S 3.3).

\subsection{Illumination by an Isotropic X-ray Source}

\begin{figure}[t]
\epsscale{1} \plottwo{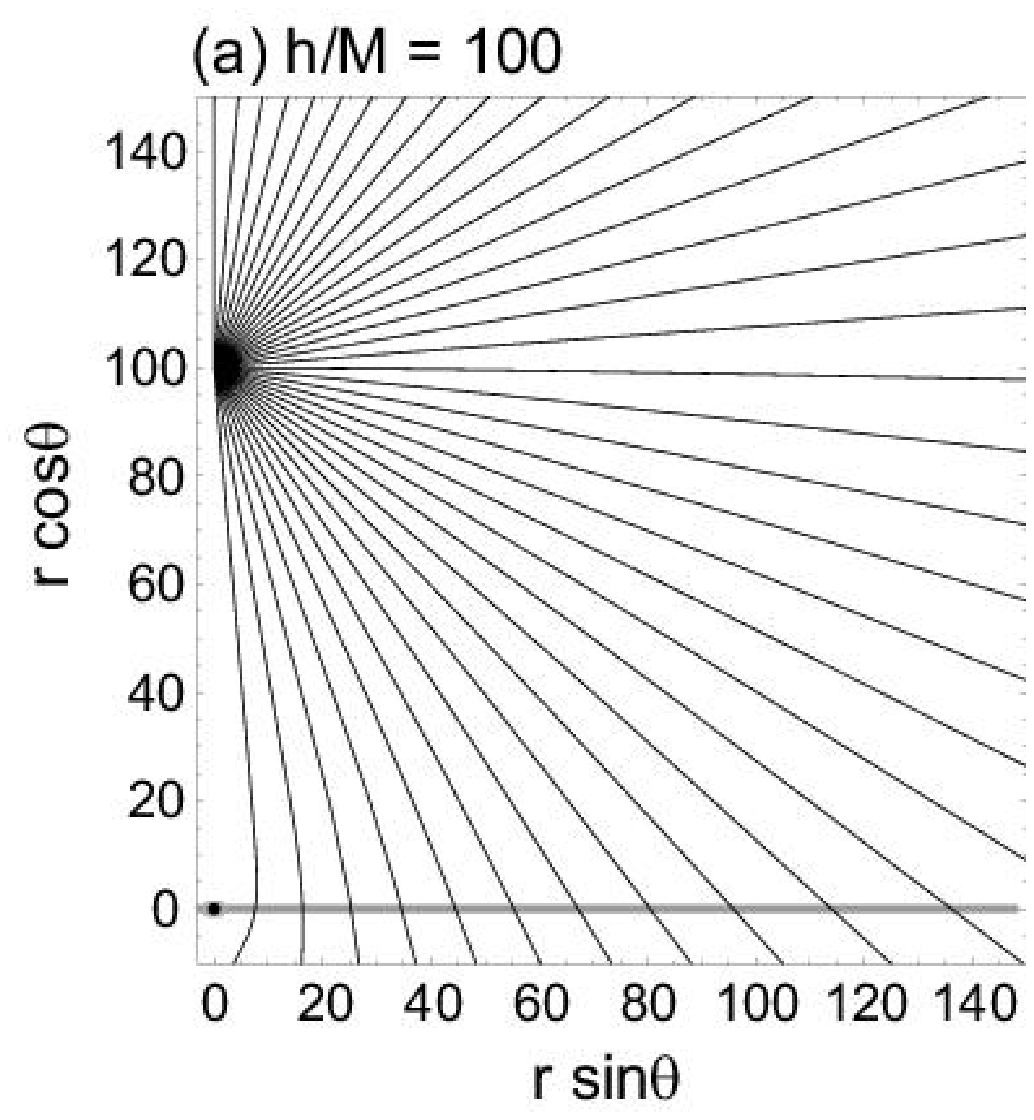}{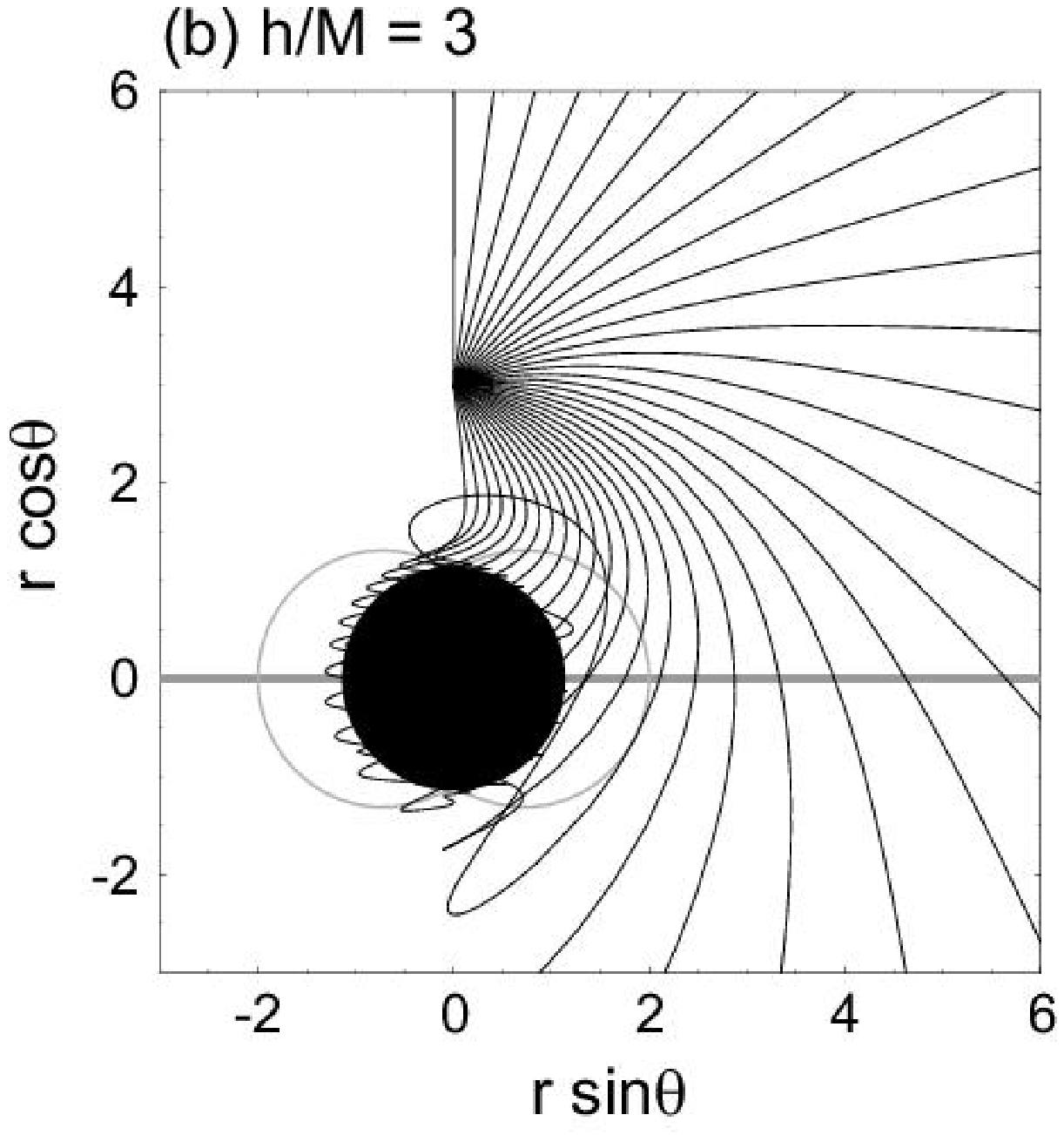} \caption{Photon trajectories
in the poloidal plane ($r,\theta$)-plane for
    (a) $h/M=100$ and (b) $h/M=3$ around a rotating black hole with $a/M=0.99$.
    An equatorial thin disk is depicted by the thick gray line at $\theta=\pi/2$
    along with a black hole at ($0,0$). The gray curve around the black hole denotes
    the static limit (ergosphere) projected onto this poloidal plane.
    Scales are different for
    presentation purposes.
\label{fig:path}}
\end{figure}

We first present our results the case of an isotropic source on the
axis of symmetry. In Figure~\ref{fig:path} we show several of the
calculated photon trajectories in the poloidal plane ($r,\theta$)
for two cases: (a) $h/M=100$ and (b) $h/M=3$. In each case the
source is isotropic locally and the photons are chosen in constant
angular size intervals of size $\Delta \psi = 8.7 \times
10^{-2}~\rm{radians} = 5\deg$ (corresponding to a total of $N=36$
orbits in the figure). Because of the axial symmetry of the problem
we only need show the trajectories of the photons emitted over $0
\le \psi \le \pi$. The value of black hole angular momentum used in
both cases is $a/M=0.99$. For an X-ray source at a height much
larger than the black hole horizon (i.e., $h/M \gg 1$) as in case
(a), the disk illumination is not very different from that of the
Minkowski geometry given above [$\propto h/(h^2+r^2)^{3/2}$] (see
e.g., Fig. \ref{fig:fit}a, also to be discussed later). On the other
hand, as the source approaches the black hole, as in case (b) of
Figure~\ref{fig:path}, the focusing influence of the geometry
becomes significant; the light rays bend toward the black hole and
enhance considerably the disk illumination in its vicinity; at the
same time, an increasing fraction of the photons are absorbed by the
black hole with the fraction approaching unity as the source
position approaches the horizon. It should be noted that in case (b)
a significant fraction of the photons are also subjected to
frame-dragging by the black hole angular momentum; this causes their
rotation in the azimuthal direction before reaching the horizon
(note the azimuthal trajectories partially seen just outside the
horizon denoted by a dark hole). However, because of the axisymmetry
of the configuration, the frame-dragging effects that influence only
the azimuthal component of the trajectory are ignorable; the
illumination depends on the photon trajectories on the poloidal
plane.

\begin{figure}[t]
\epsscale{0.7} \plotone{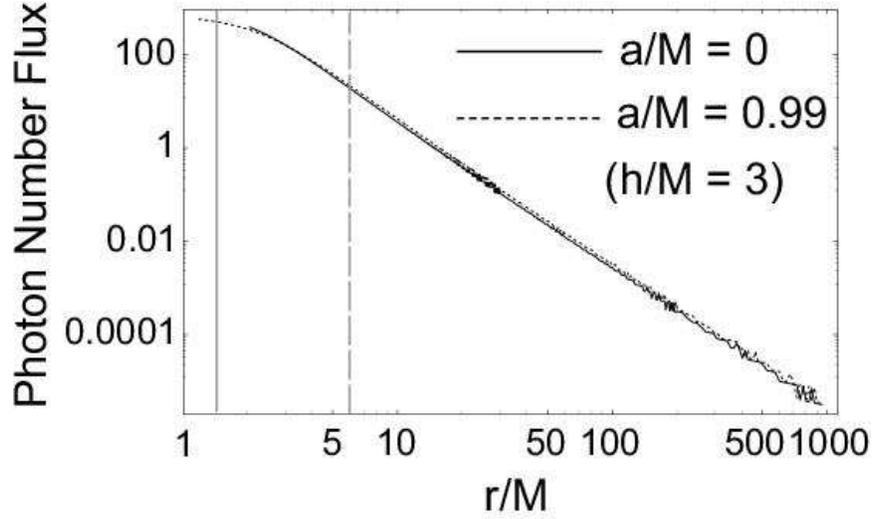} \caption{Photon number illumination
laws as a function of the normalized disk radius $r/M$ for a \sw and
a Kerr black hole with the values of $a$ noted in the figure. In
both cases the source height has been $h/M=3$. Radius of the ISCO is
denoted by vertical dashed line for $a/M=0$ and solid line for
$a/M=0.99$. The illumination in each case has been computed to $r =
r_{\rm{H}}$. } \label{fig:spin}
\end{figure}

The main effects of the black hole angular momentum $a$, as far as
our current considerations go, mainly lie in the reduction of the
sizes of the horizon $r_{\rm{H}}$ and, most importantly, that of
ISCO, which allows the illumination of regions of the disk with
sufficiently large redshift, a necessary condition for producing the
very broad Fe lines observed in MCG --6-30-15. While the photon
orbits in the poloidal plane can change significantly with
increasing $a$, the illumination law depends mainly on the height of
the source above the disk $h/M$ and it is largely independent of the
value of $a$ (this is true for $h/M \gsim 3$; for smaller values of
this parameter, the illumination laws do differ substantially). We
have verified the above statement by direct orbit computation:
Figure \ref{fig:spin} exhibits the illumination profiles of
accretion disks around a \sw (solid) and a Kerr black hole of
$a/M=0.99$ (dotted curve) respectively by a source at $h/M=3$. Even
for such a small height, the illumination patterns are almost
identical, except for the fact that a high value of $a$ allows
smaller value of the horizon $r_{\rm{H}}$ and of ISCO. Because of
this the illumination profiles obtained for a Kerr black hole are
also appropriate for a black hole with any value of $a$, provided
that the illumination is restricted to the proper value of ISCO.


\begin{figure}[t]
\epsscale{0.7} \plotone{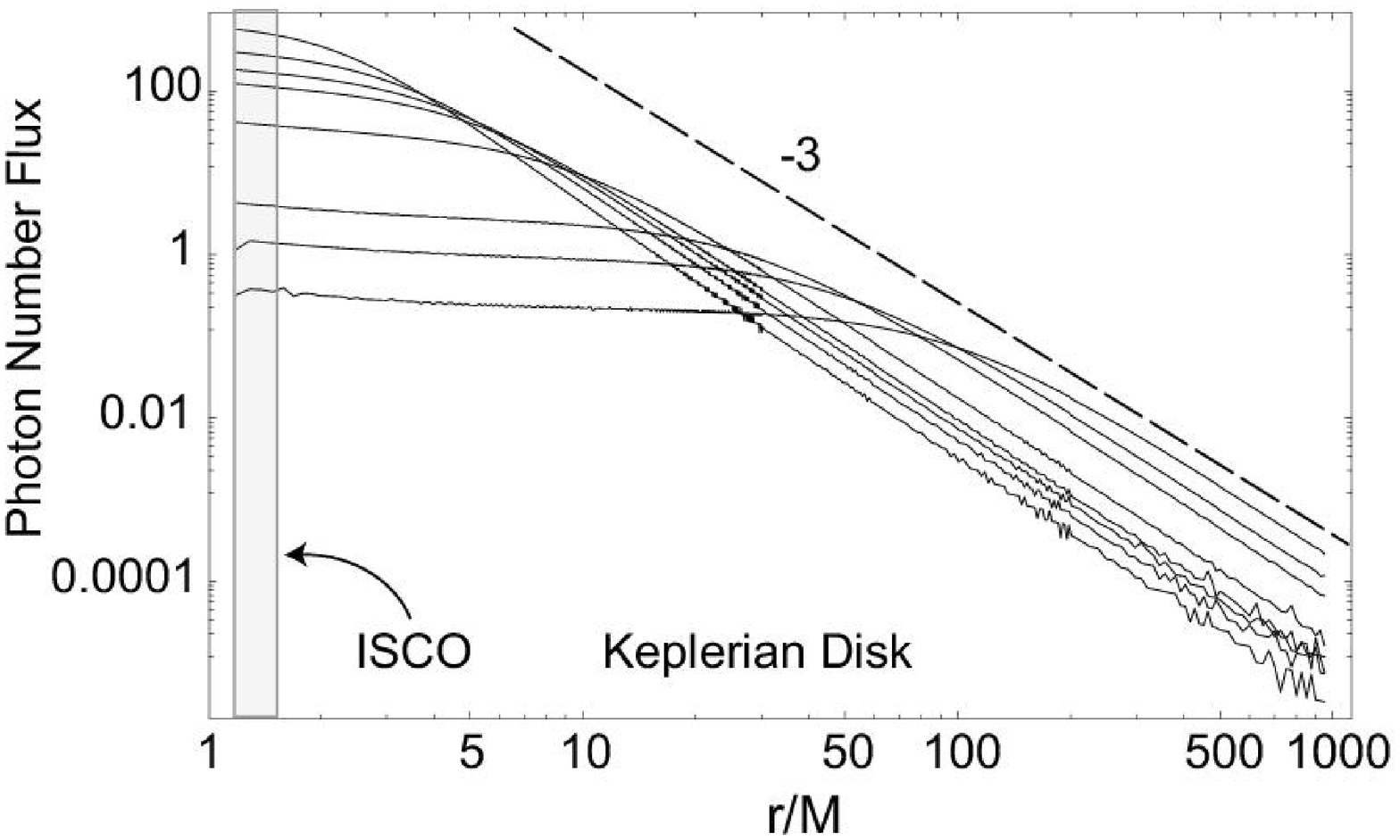} \caption{Radial illumination
profiles of the photon number flux at the disk plane for an
    isotropic flux of 40,000 photons at the source. We take $a/M=0.99$.
    The corresponding source heights are $h/M=100, 50, 30, 10, 6, 5, 4$
    and $3$,
    from bottom to top at the left side of the figure. Dashed line
    denotes a slope of $-3$ for power-law form ($r^{-3}$). Plunging region inside ISCO
    is shown by a shaded box.} \label{fig:IL-1}
\end{figure} 

By computing a sufficiently large number of orbits in a Kerr
geometry with $a/M=0.99$, similar to those appearing in
Figure~\ref{fig:path}, and keeping track of their normalization
(proportional to sin$\,\psi$) we have calculated the radial
distribution of the illumination by collecting the photons
intercepting the disk plane within a given range in radius. The
results for different values of the parameter $h/M$ are shown in
Figure~\ref{fig:IL-1}; the values of this parameter used are:
$h/M=100, 50, 30, 10, 6, 5, 4$ and $3$. The inner radius is set to
$r_{\rm{H}}/M=1.15$ (for $a/M=0.99$) and the outer radius is set to
$r_{\rm{out}}/M=1000$. In all cases, the illumination approaches
asymptotically (i.e., for $r \gg h$) the functional form appropriate
for Minkowski space [$\propto h/(r^2+h^2)^{3/2} \sim r^{-3}$ for $r
\gg h $]; this result is not unexpected as the photons that
intercept the outer reaches of the disk have been emitted almost
horizontally at the source, directions at which the influence of the
focusing effects of the black hole are minimal for the values of
$h/M$ considered here. The normalization of the illumination in this
regime of the radius $r$ is very close to that of Minkowski space
too except for the case $h/M=3$ in which the gravitational effects
``pull" the photon trajectories toward the black hole reducing the
normalization below its flat space value (see also
Fig.~\ref{fig:fit}d to be explained later). At values of $r \simeq
h$ the profile ``levels-off" to a slope which approaches that of
flat space ($\propto r^0$). The effects of gravitational focusing
begin to become noticeable for $r/M \lsim 7$ (even for very large
values of $h/M$), as shown in Figure~\ref{fig:fit}a; their
importance increases with decreasing values of $h/M$, as expected
and shown in Figure~\ref{fig:fit}a-\ref{fig:fit}d, where the actual
illumination is compared to that corresponding to the Minkowski
geometry.


The numerical accuracy of our scheme can be assessed by comparing
the normalization of the flux of our calculations for large values
of $h/M$ with the analytic result of the flat geometry, which for
$r\ll h$ is $F_{\rm{flat}} \simeq N_{\rm tot}/ (4 \pi h^2)$, where
$N_{\rm tot}$ is the total number of photons emitted by an isotropic
source. Our computation proceeds by launching photons starting at
$\psi =0$ and incrementing the launch angle by $\Delta \psi = 5
\times 10^{-5}$ radians. The total number of orbits we compute is
then $N = \pi /\Delta \psi$; since each photon is weighted by the
sine of the emission angle sin$\psi$ (to take into account the
number of photons emitted in all azimuths) the corresponding total
number of photons emitted by an isotropic source, $N_{\rm tot}$, is
related to the number of orbits computed, $N$, by the relation
\begin{equation}
N_{\rm tot} = \sum_{i=0}^{N} {\rm sin}(i\Delta \psi) =\sum_{i=0}^{N} {\rm sin}
\left(\frac{i \pi}{N} \right) = \cot\left(\frac{\pi}{2N}\right) \simeq \frac{\pi}{\Delta
\psi}\frac{2}{\pi}~~({\rm for~} N \gg 1).
\end{equation}
Using this value for $N_{\rm tot}$ and assuming $h/M = 100$ (to
provide adequate approximation to Minkowski geometry) one obtains
for the flux at $M \ll  r \ll h$ (using $\Delta \psi = 5 \times
10^{-5}$) the value $F_{\rm{flat}} = 1/ \pi$, in excellent agreement
with the value shown in Figure~\ref{fig:IL-1} for the specific value
of $h$. Our scheme therefore, provides the correct shape and
absolute normalization of the illumination at the appropriate limits
and we are confident that it provides also the absolute
normalization and illumination shape in the regime in which the
focusing effects of the geometry are important.


\begin{figure}[t]
\epsscale{1.1} \plottwo{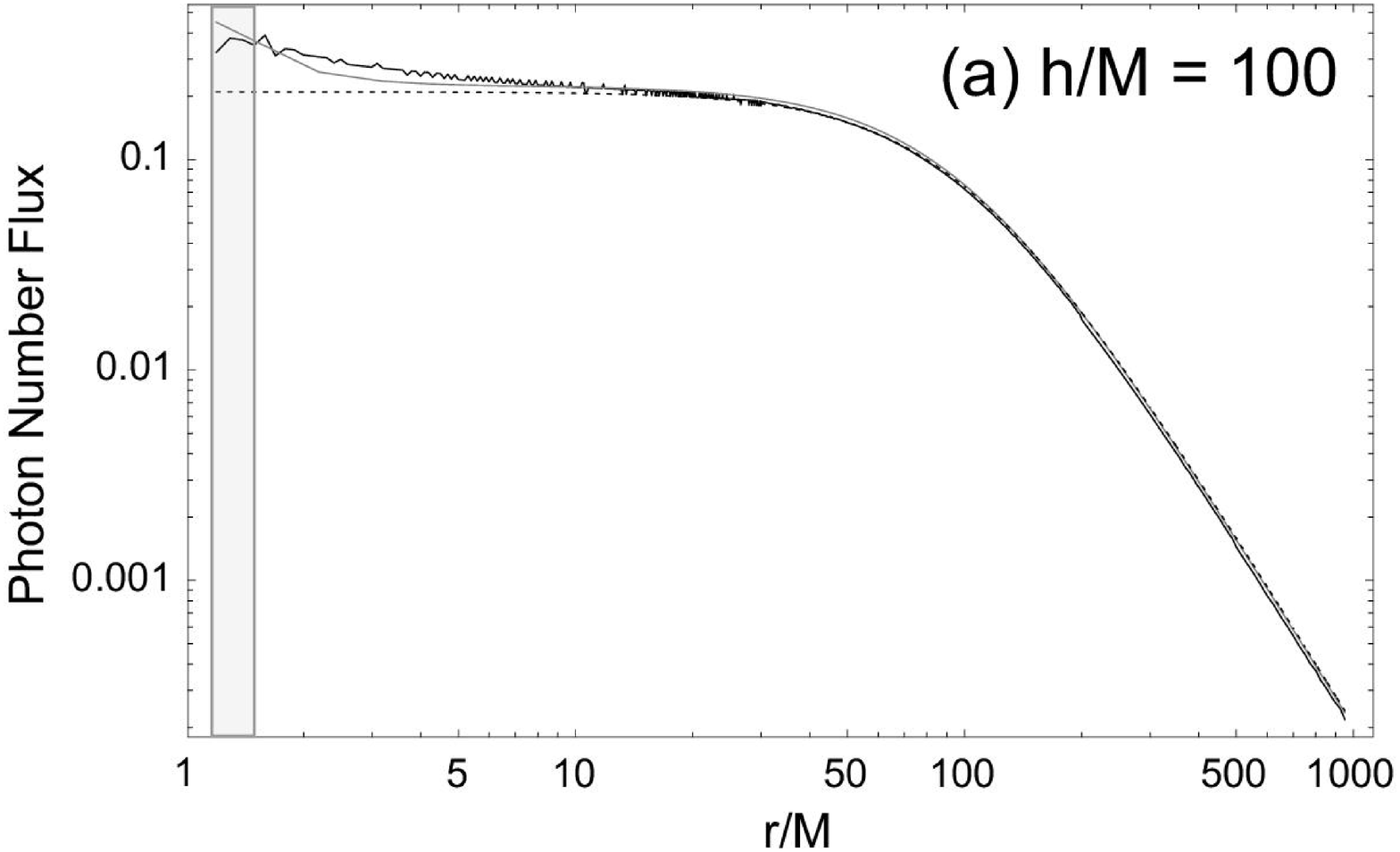}{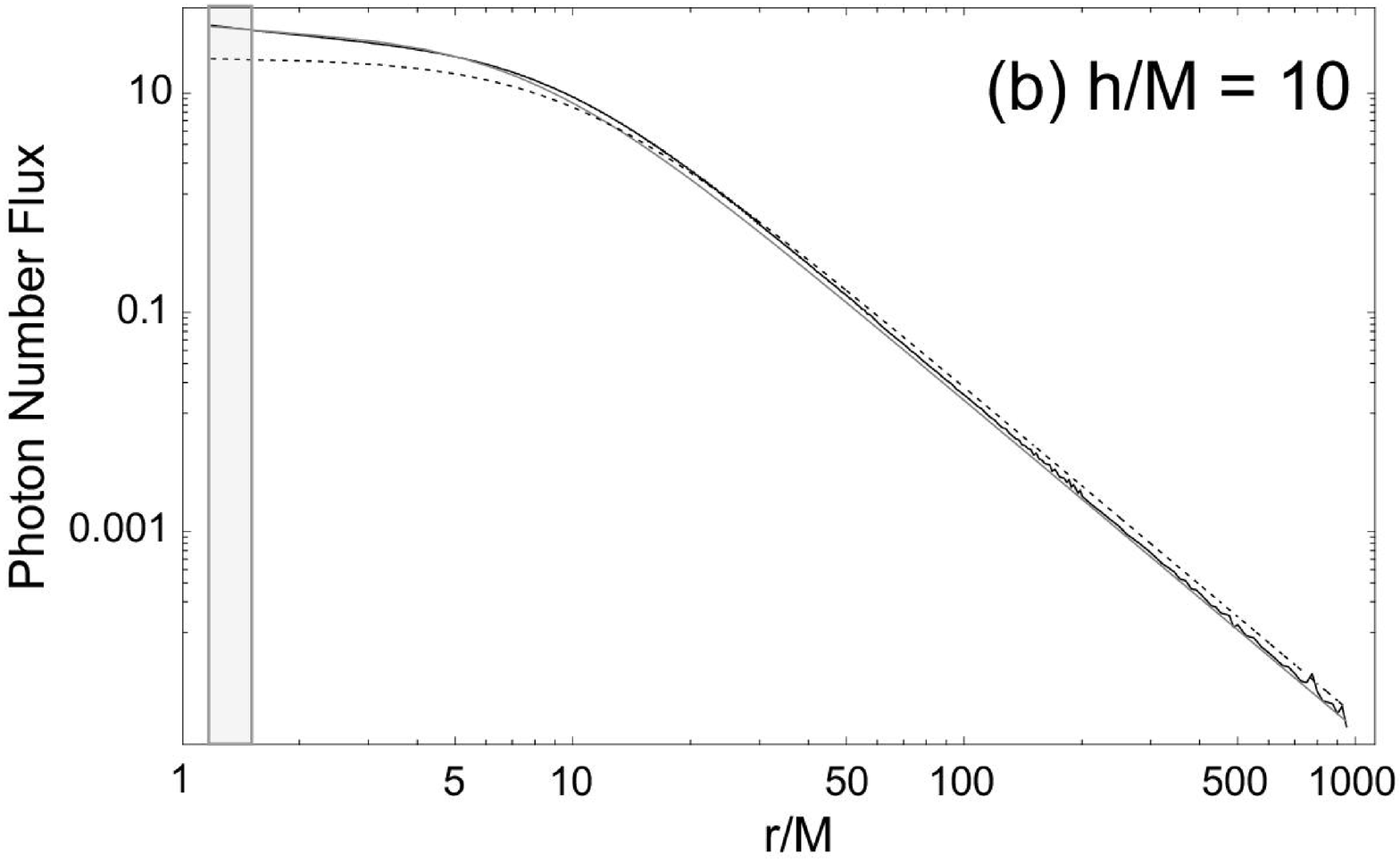}\plottwo{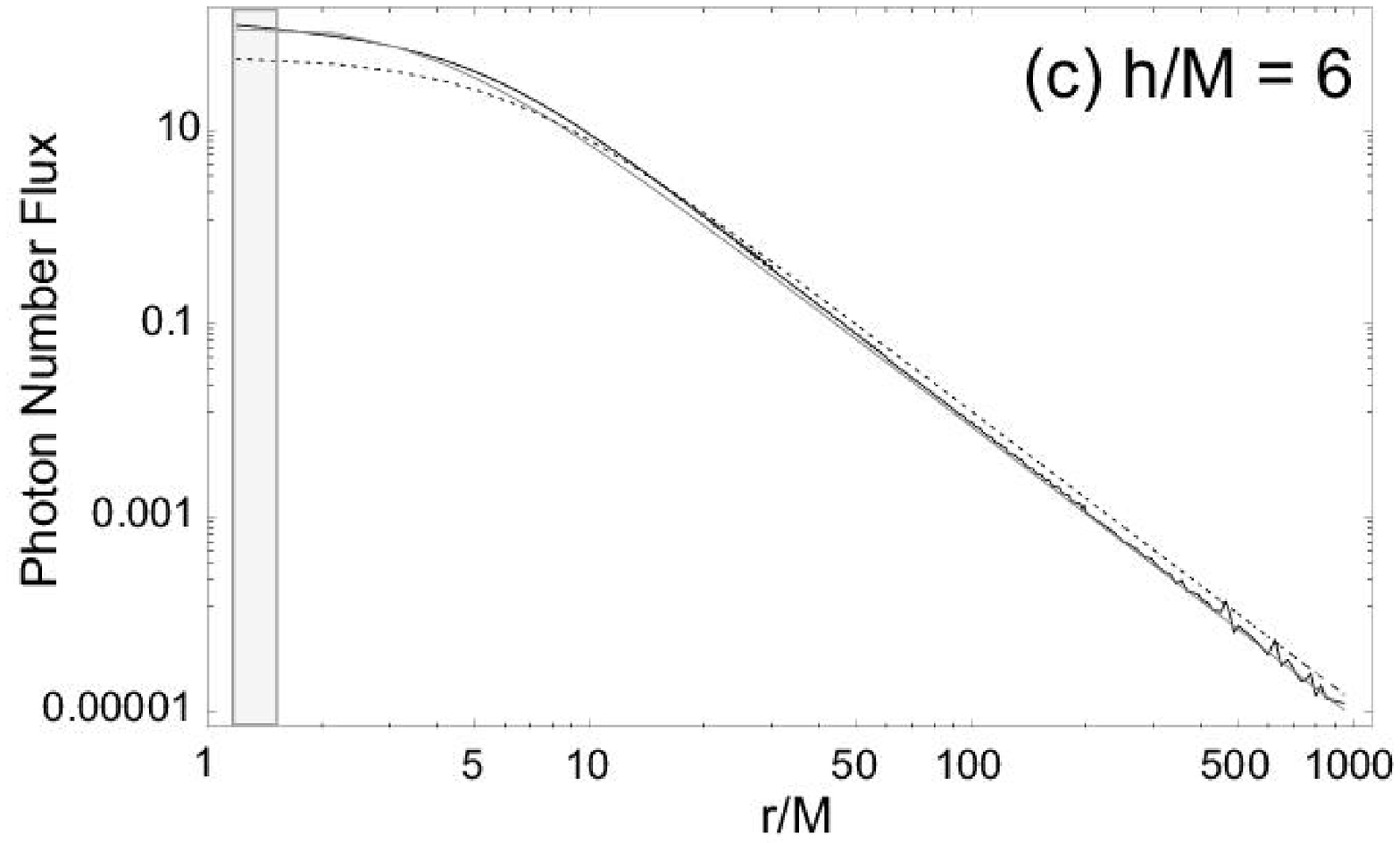}{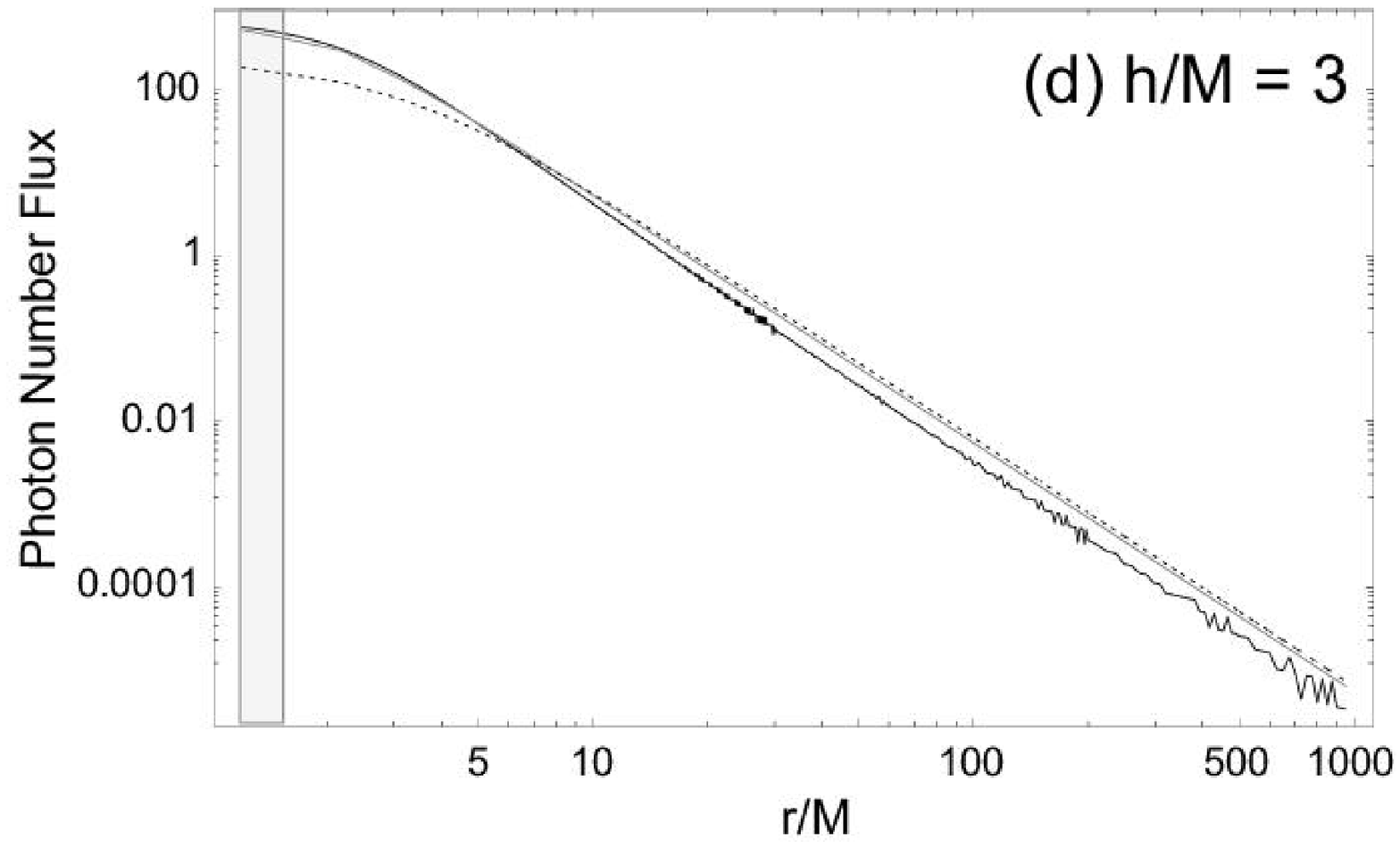}
\caption{The same illumination profiles (solid dark curves) as in
Figure~\ref{fig:IL-1}
    for (a) h/M=100, (b) 10,
    (c) 6, and (d) 3 along with our best-fitted analytic function (solid gray curves) expressed by
    equation~(\ref{eq:fit}). Dotted curves represent the corresponding profiles
    in Minkowski space. \label{fig:fit}}
\end{figure}

The numerically obtained illumination profiles of
Figure~\ref{fig:IL-1} were fit with an empirical function of the
following form (in the parameter interval noted)
%
\begin{eqnarray}
F(r,h;K) = K \times \left\{ \begin{array}{llr}
F_1(r,h) & ~~~{\rm for}~~~ 6 < h/M \le 100 ~, & r/M > 1.15  \\
F_2(r,h) & ~~~{\rm for}~~~ 3 \le h/M \le 6 ~,&  r/M > 1.15\\
\end{array} \right. \label{eq:fit}
\end{eqnarray}
where $K$ is the normalization and the functions
$F_{1,2}(r,h)$ are given by
\begin{eqnarray}
F_1(r,h) &=& -\frac{1.026 \times 10^{-5}
(h-152.9)(h^2-158.7h+6569)}{r^3} \nonumber \\ & & - \frac{3.364
\times 10^{-2} (h-348.7)(h+11.98)(h+139.4)}{\left\{h^2+\left(1+1/h
\right)^6 r^2 \right\}^{3/2}}  \ ,
\\ \nonumber \\
F_2(r,h) &=& \frac{79.56 (h-6.250)(h^2-9.763h+25.19)}{r^3} \nonumber
\\ & & - \frac{3921
(h-7.364)(h^2-8.556h+21.92)}{\left\{h^2+\left(1+1/h \right)^6 r^2
\right\}^{3/2}} \ ,
\end{eqnarray}
where $h$ and $r$ are measured in units of $M$; the numerical constants
that provide the normalization of each of the fractions of the fitting
formulae were chosen to give the correct normalization for a total number
of photons (by an isotropic source) equal to $N_{\rm tot} = 2/\Delta \psi
= 40,000$, i.e. the number appropriate  to our own simulations.

In Figure~\ref{fig:fit} the fits provided by this formula (gray)
along with the numerically computed profiles (solid dark) and the
corresponding Minkowski geometry profile (dotted curves) are shown
for (a) $h/M=100$, (b) $h/M=10$, (c) $h/M=6$, and (d) $h/M=3$. It is
apparent that the analytic formulae provide very good fits to the
computed curves both in shape and normalization, especially at $r
\lsim h$, the region crucial for the correct determination of the
relativistic iron line shapes.


It should be noted at this point the above calculations provide the
fraction of the total number of photons (number flux) emitted by the
source that are intercepted between $r$ and $r + dr$ on the disk as measured
in Boyer-Lindquist coordinates $(t,~ r,~ \theta,~ \phi)$ by an
observer at rest with respect to the photon source. However,
conventional calculations of Fe line profile involve the rate of Fe
line photon production as a function of the coordinate $r$, measured in the
same coordinates by an observer in relative motion with respect to
the photon source with the local Keplerian velocity. Modulo an efficiency factor
for the conversion of continuum into line photons, the transformation of
the rates between these two Lorentz frames involves powers of the
factor
\begin{equation}
{\cal D} \equiv \frac{u^{\mu}_{\rm r} p_{\mu}} {u^{\mu}_{\rm e} p_{\mu}} \ ,
\label{eq:Dfactor}
\end{equation}
where $u^{\mu}_{\rm e},
~ u^{\mu}_{\rm r}$ are respectively the four-velocities of the
emitter and receiver of radiation and $p_{\mu}$ the photon
four-momenta. The expression of this factor for a source at rest in
the Boyer-Lindquist frame and a receiver at rest with the frame
moving with the local (disk) Keplerian velocity is computed in
Appendix B. The analytic expressions given by
equation~(\ref{eq:fit}) can be transformed to this frame by
multiplying by ${\cal D}^{1+\alpha}$ where $\alpha$ is the spectral
index of the power-law X-ray source (assumed to have the form
$F_{\nu} \propto \nu^{-\alpha}$ in the local source frame): one
power of ${\cal D}$ indicates the change in the rate photons are
intercepted by the comoving observer in the fluid/disk frame, while
the factor ${\cal D}^{\alpha}$ accounts for the spectral shift in
the photon energy as perceived by the same comoving observer.

\begin{figure}[t]
\epsscale{1.1} \plottwo{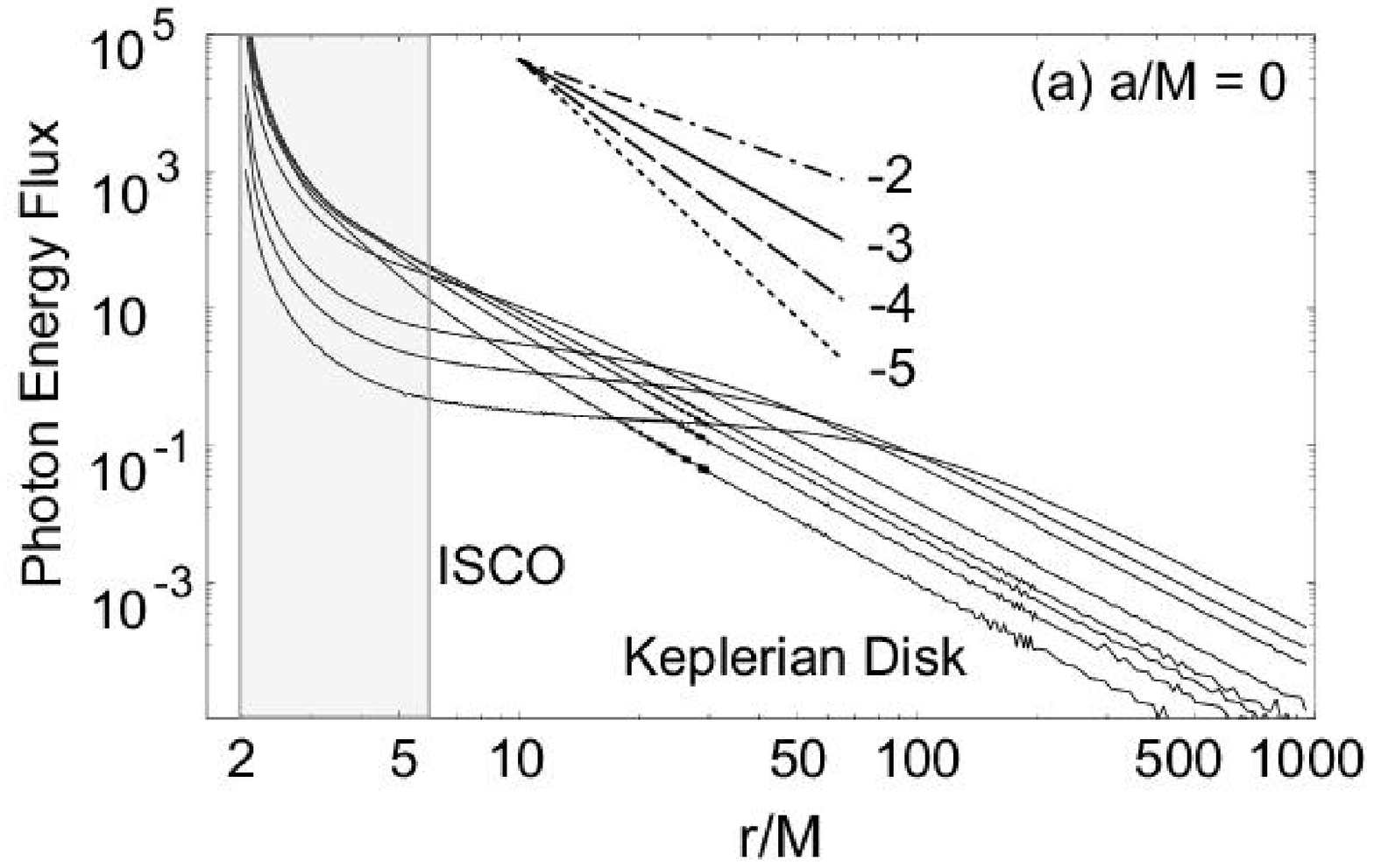}{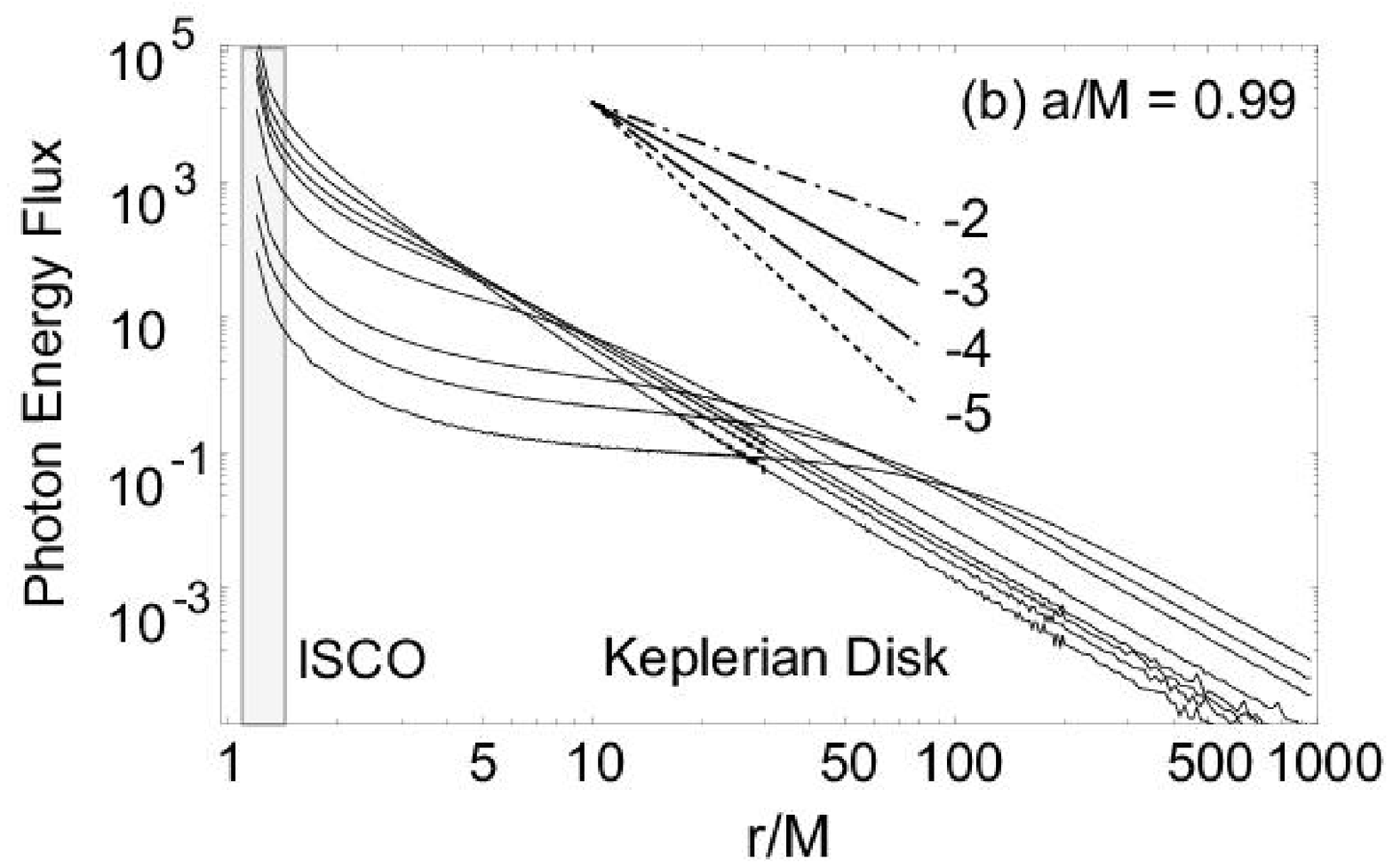} \caption{Radial
illumination profiles, corresponding to Figure~\ref{fig:IL-1}, of
the photon energy flux in the comoving fluid frame. Various slopes
are shown as a reference. \label{fig:IL-2}}
\end{figure}

In Figure~\ref{fig:IL-2} we present the results of this section
modified by the above factor (photon energy flux) for a \sw and a
Kerr ($a/M = 0.99$) geometry; i.e. $F(r,h;K) {\cal{D}}^{1+\alpha}$.
We assume a Keplerian motion outside the ISCO while a free-fall
motion in the plunging region. One sees that in both cases there is
a considerable increase in the line production rate for sufficiently
small radii. Since the slope of illumination in photon energy flux
is always steeper than that of the photon number flux by the factor
of ${\cal{D}}^{1+\alpha}$, the illumination profile of the photon
energy flux exhibit two characteristics: the slope of the outer disk
zone follows that of flat space ($q \sim 3$), while in the inner
disk zone (outside the ISCO) the slope becomes (somewhat) steeper
($3 \lesssim q \lesssim 4$). Inside the ISCO the slope becomes
extremely steep simply because of the factor ${\cal{D}}^{1+\alpha}$
. Assuming that the disks terminate at the ISCO and that there is
little emission interior to this radius (as shown in Fig.~3 of
Reynolds \& Begelman 1997 the matter is expected to be highly
ionized in this region, except in cases of very low efficiency), our
results indicate that in the case of a \sw geometry the
amplification of the photon flux due to the position and motion of
the absorbing plasma on the disk makes only a small change to the
results of Figure~\ref{fig:IL-1}. Howerver, the situation is
different in the Kerr geometry examined($a/M=0.99$), where the above
effect becomes significant at regions of the disk outside the ISCO .



\begin{figure}[t]
\epsscale{0.5} \plotone{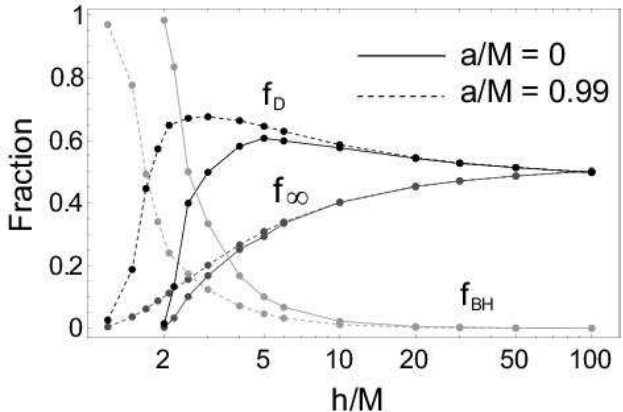} \caption{The fractions of the
photons emitted by an isotropic source that reach
    the disk plane $f_{\rm{D}}$, escape to infinity $f_\infty$, and
    lost through the horizon $f_{\rm{BH}}$,
    as a function of $h$. We consider $a/M=0$ (solid) and $a/M=0.99$ (dotted curves) cases.
    The cut-off radius is chosen to be
    $R/M=2000$. }
\label{fig:fraction1}
\end{figure} 

A different aspect of the illumination problem concerns the fraction
of the photons of an isotropic source that is intercepted by the
accretion disk plane. For a point-like source at height $h$ on the
symmetry axis in Minkowski geometry, the total number of photons
intercepted in the region $r_{\rm{H}} < r \le R$ of the disk is
given by
\begin{equation}
N_{\rm{flat}} = N_{\rm tot}\, \frac{1}{\pi} \,
\left[\cos^{-1}(h/\sqrt{h^2+R^2})-\cos^{-1}(h/\sqrt{h^2+1})\right] \
. \label{eq:counts}
\end{equation}
This is the fraction of the total number of photons, $N_{\rm tot}$,
emitted within the solid angle subtended by the disk (excluding the
black hole horizon, taken in this case to have the value
$r_{\rm{H}}/M = 1.15  \simeq 1$). For $R \gg h \gg 1$ this attains
the value $N_{\rm tot}/2$, the remaining photons escaping to
infinity.  The focusing effects of the geometry modify significantly
the fraction of photons intercepted by the disk, escape to infinity
or are absorbed by the black hole as the source approaches the
horizon. The procedure we have used in \S 2 can be used to compute
these fractions which are of value in estimating the effects of the
geometry on the ratio of line flux to continuum, i.e., the line
equivalent width (EW) or the strength of the Compton reflection
feature.
Figure~\ref{fig:fraction1} shows the percentage of the photons of an
isotropic source that reach the disk plane (out to a distance $R/M =
2000$), $f_{\rm{D}}$, that of photons escaping to infinity,
$f_{\infty}$, as well as the fraction absorbed by the black hole,
$f_{\rm{BH}}$, as a function of the source height $h$ for $a/M=0$
(solid) and $a/M=0.99$ (dotted curves). For large $h$ ($h/M \gsim
50$) $f_{\rm{D}}, ~f_{\infty} \rightarrow 1/2$ while $f_{\rm{BH}}
\rightarrow 0$, as expected. As the height of the source decreases,
$f_{\infty}$ decreases monotonically to zero (for $h \rightarrow
r_{\rm{H}}$); however, $f_{\rm{D}}$ increases initially reaching a
maximum of $\simeq 70\%$ for $h/M \simeq 3$ ($\simeq 60\%$ at $h/M=
5$ for a \sw black hole) before it declines to zero as the source
approaches the black hole horizon. Last, the fraction $f_{\rm{BH}}$
does have the qualitative behavior expected, being negligible for
large $h$, reaching unity as $h \rightarrow r_{\rm{H}}$. The
qualitative behavior of these quantities does not change as one
moves from the case of an extreme Kerr geometry (dotted) to that of
a \sw black hole (solid curves). However, the maximum value of
$f_{\rm{D}}$ is slightly smaller and it is achieved for a slightly
larger value of $h$. It is of interest to note that the values of
$f_{\rm{BH}},~f_{\rm{D}},~f_{\infty}$ are very similar to those
presented by \citet{Miniutti03}, who considered a different source
geometry (a ring-like source in an extreme Kerr geometry of radius
$r/M=2$ located at a height $h$) and a different computational
approach. It appears therefore, that this represents a generic
behavior for illumination not only by point-like but also by
extended X-ray sources provided that they are concentrated near the
symmetry axis of the configuration (rather than, e.g., near the
equator).

\subsection{Illumination by an Anisotropic X-ray Source}

While the formulae given above provide very good approximations to
disk illumination  by an isotropic source, we have chosen to produce
also the illumination by an anisotropic emission of photons,
appropriate for cases that the observations demand illumination more
concentrated near the disk inner edge than that of an isotropic
source. Clearly this is an arbitrary choice for if one moves away
from the isotropic assumption there is little constraint as to how
much anisotropy should be used. Nonetheless, in order to provide an
algorithm as well as a gauge of the source anisotropy with a certain
theoretical basis we have decided to Lorentz-boost a locally
isotropic source distribution along the symmetry axis and toward the
black hole by a constant velocity $v$. We assume the emission to be
instantaneous, i.e., that the photons are emitted before the source
has moved substantially to alter its height $h$.

The methodology followed is similar to that used earlier: a value of
the local angle $\psi^{\prime}$ is chosen uniformly as before to
provide the photon distribution at the rest frame of the emitter. We
then compute the corresponding angle $\psi$ in the local rest frame
which is related to $\psi^{\prime}$ by the usual angle
transformation
\begin{equation}
{\rm tan} \psi = \frac{{\rm sin} \psi^{\prime}}{\gamma({\rm
cos}\psi^{\prime} - \beta)} \ ,
\end{equation}
where $\beta \equiv v/c$ and $\gamma \equiv 1/\sqrt{1-\beta^2}$.
This value of the angle is then used along with equation
(\ref{angle2}) to obtain the values of $b$ and $\dot r$ used in the
integration of the photon orbits. The factor sin$\psi$ of this
latter angle $\psi$ is again used to provide the correct
normalization of the photons illuminating the disk.

As expected, the anisotropy of emission, measured by the velocity of
the corresponding relativistic motion, provides a higher
concentration of the available photons towards the inner edge of the
disk than the geometry alone for a source approaching the disk
($\beta>0$) and correspondingly less concentrated for a source
moving away from the disk ($\beta<0$). At the same time, depending
on the anisotropy and the height of the source, a large fraction of
the photons can be lost through the black hole horizon.

\begin{figure}[t]
\epsscale{1} \plottwo{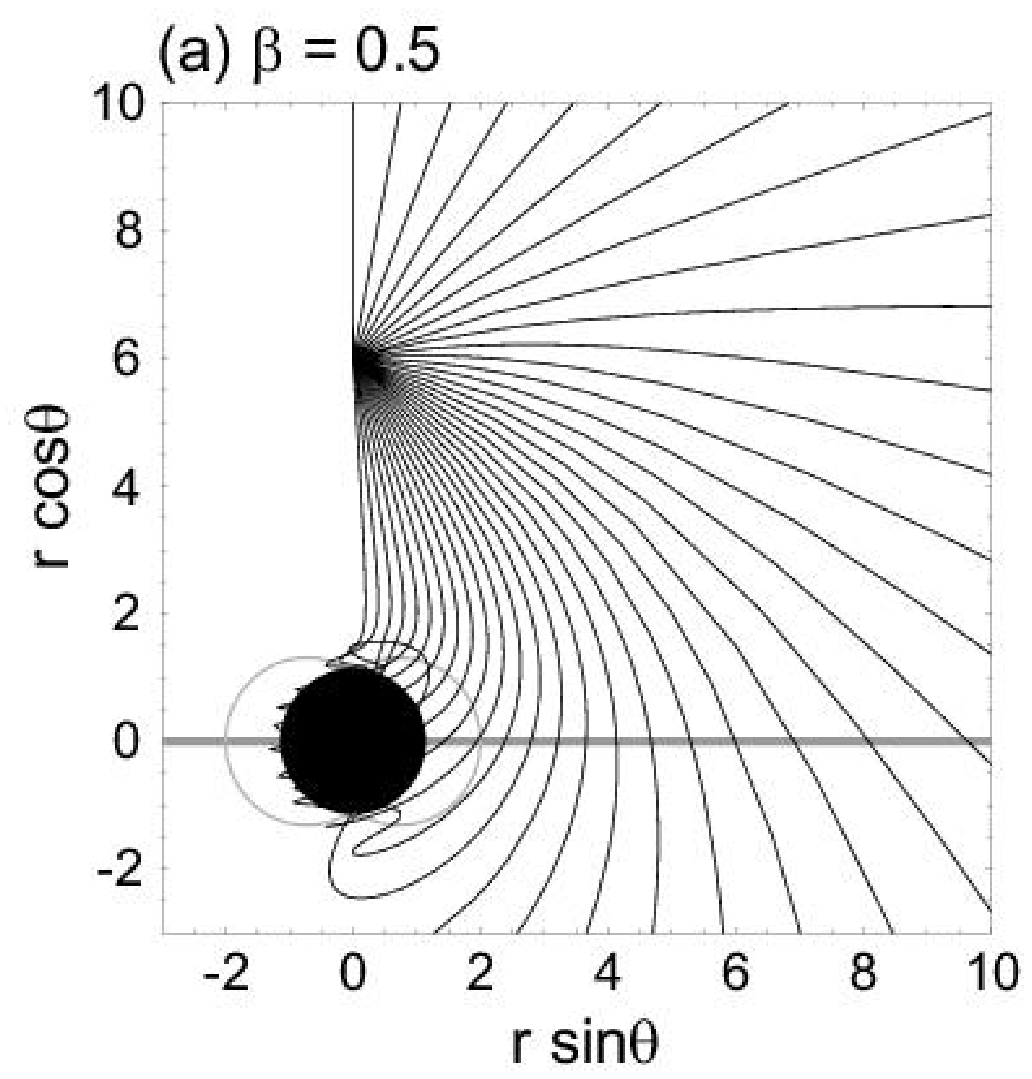}{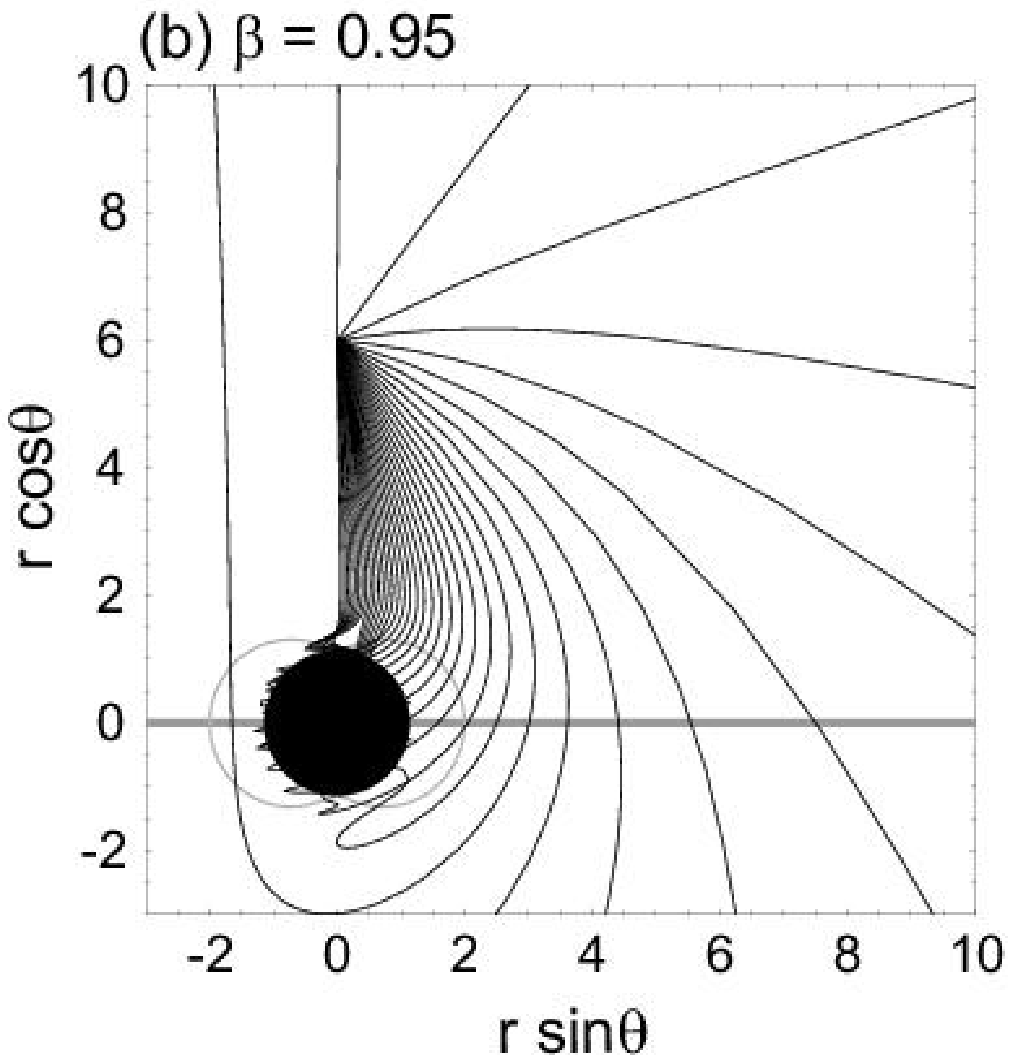} \caption{The photon
trajectories in the
    poloidal plane for an anisotropic source at $h/M=6$ with
    (a) $\beta=0.5$ and (b) 0.95. } \label{fig:beaming1}
\end{figure}

\begin{figure}[t]
\epsscale{1} \plottwo{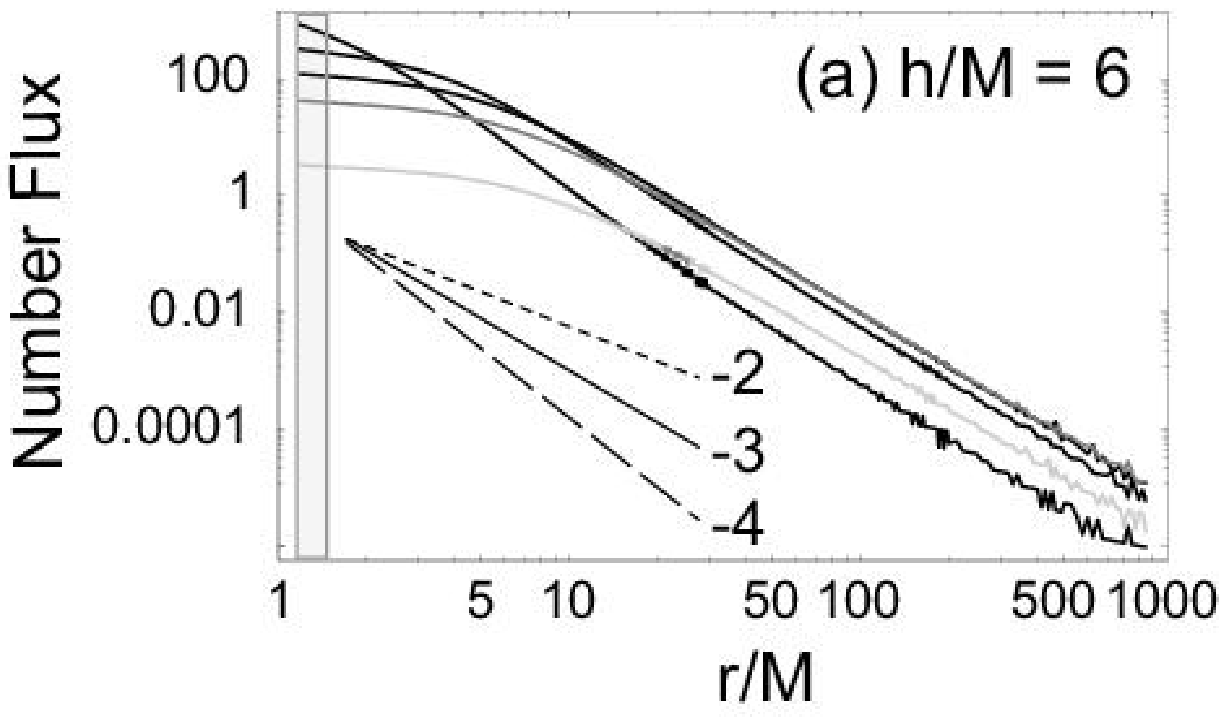}{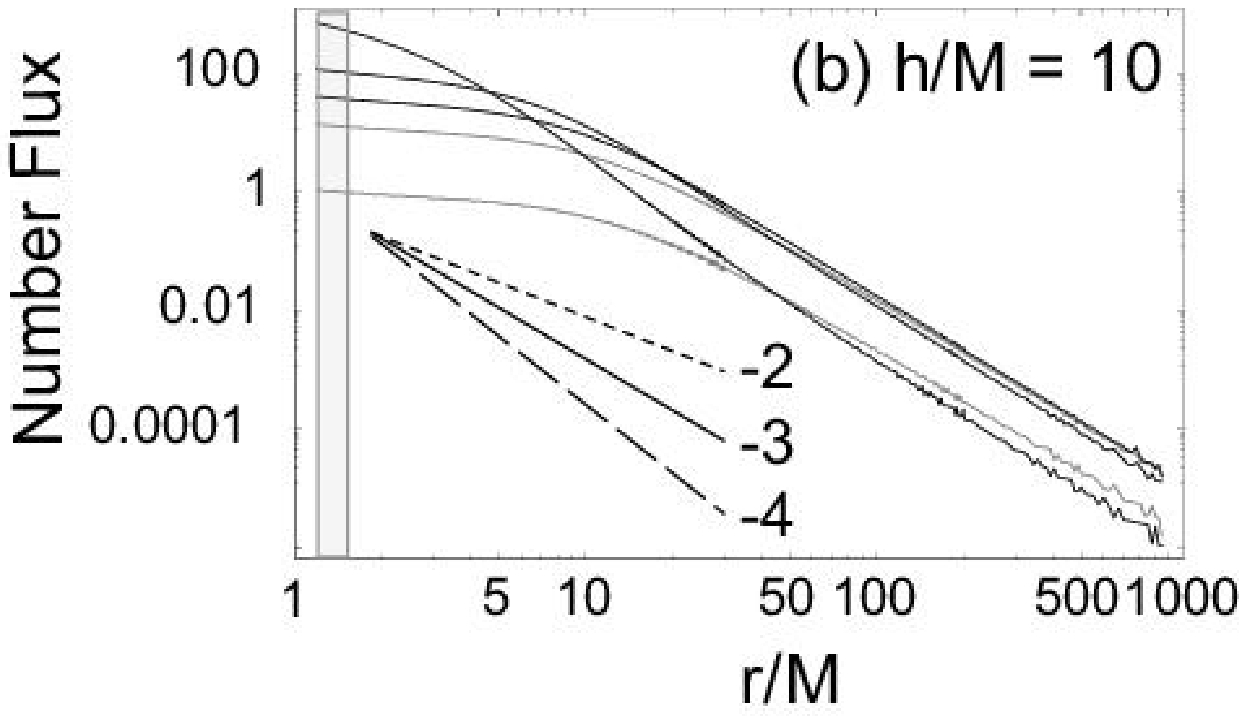} \caption{The illumination
profiles for $\beta = -0.95$ (gray), $-0.5$ (gray), $0, 0.5, 0.95$
from bottom to
    top at the left side of the figure for (a) $h/M=6$ and (b) $10$.
    Three lines represent the slopes
    for $-2$ (dotted), $-3$ (solid) and $-4$ (dashed lines) as a reference. }
\label{fig:beaming2}
\end{figure}

Figure~\ref{fig:beaming1} shows the photon trajectories by an
anisotropic source (moving toward the disk) of anisotropy
corresponding to instantaneous velocities (a) $\beta=0.5$ and (b)
$\beta = 0.95$ at a height $h/M=6$ illuminating a disk around a
black hole of $a/M=0.99$. Figure~\ref{fig:beaming2} displays the
corresponding disk illumination of an anisotropic source for two
different values of the source height, namely (a) $h/M=6$ and (b)
$h/M=10$. For each value of $h/M$ we consider for comparison source
anisotropies corresponding to $\beta=-0.95,\; -0.5,~0,~0.5,~0.95$,
with the minus sign indicating motion away from the accretion disk.
As expected, the anisotropy increases disk illumination for $r
\lesssim h$ for sources approaching the disk, while decreasing it
for sources receding from the disk; its slope becomes
correspondingly steeper for the higher value of the anisotropy
parameter $\beta = v/c$. As a result, the profile can deviate
greatly from that of an isotropic source. Qualitatively, the
increase in $\beta$ mimics a decrease in $h$ to the point that ${\rm
cos} \, \psi \simeq 1/\gamma = \sqrt{1 - \beta^2}$, that is to the
point that the Lorentz boosting solid angle ($\propto 1/\gamma^2$)
is roughly equal to the solid angle subtended at the source position
by the black hole horizon; at this value of $\beta$ the illumination
can be approximated by a power-law $r^{-3}$ for the entire $r-$range
as shown in Figure~\ref{fig:beaming2}; correspondingly a negative
value of $\beta$ (a receding source) provides an illumination
profile equivalent to a source at a higher value of $h$. We have
checked by direct computation that further increase in the
anisotropy through an increase in $\beta$ (or $\gamma$) preserves
the $r^{-3}$ functional form of the illumination (at all distances)
but decreases its normalization in proportion to $\gamma^{-2}$ as
might be expected on the basis of qualitative considerations; by
contrast increasing of $|\beta|$ for a receding source, increases
indefinitely the effective height of the source $h$. Cases that
demand an illumination law steeper than $r^{-3}$ (in number flux)
must therefore involve assumptions about the source and its geometry
different from those hitherto considered (see \S 4). On the other
hand, energy flux profile can become steeper than $\sim r^{-3}$ by
the factor of ${\cal{D}}^{1+\alpha}$ in energy flux, as explained
earlier. In this respect it is worth considering that a source that
could reverse its axial motion ($\pm \beta$) would lead to
substantial change in the number of photons intercepting the
accretion disk and the flux of the Fe line, even for moderate values
of $|\beta| ~(\simeq 0.5)$.

\begin{figure}[t]
\epsscale{0.5} \plotone{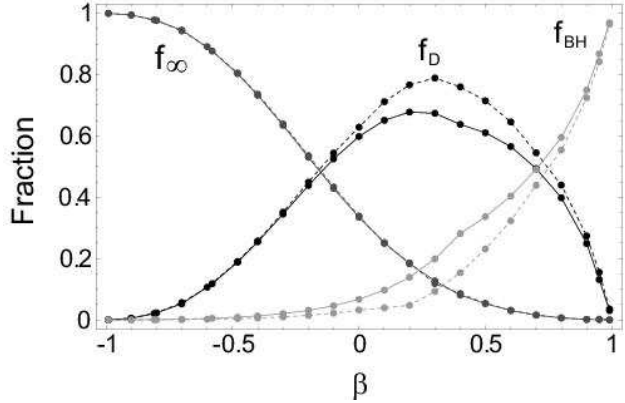} \caption{Same as
Fig.~\ref{fig:fraction1} but for anisotropic cases. A source
approaching the black hole has $\beta>0$. The source height is set
to be $h/M=6$. } \label{fig:fraction2}
\end{figure} 

For the reasons discussed in the previous subsection we have also
computed for the anisotropic illumination case the fractions
$f_{\rm{BH}},~f_{\rm{D}},~ f_{\infty}$ of the photons intercepted by
the black hole, the accretion disk and escaping to infinity,
applying the same methodology used in \S 3.1. These results are
shown in Figure~\ref{fig:fraction2} for a source at a fixed height
$h/M=6$ as a function of the anisotropy parameter $\beta = v/c$ for
$a/M=0$ (solid) and $a/M=0.99$ (dotted curves). For $\beta = 0$
these fractions are those corresponding to $h/M=6$ of
Figure~\ref{fig:fraction1}, as they should. As shown in this figure
the maximum (integrated) disk illumination, $f_{\rm{D}}$, is
obtained for $\beta \simeq 0.3$. Also as expected, $f_{\infty}$
decreases with increasing $\beta$ while $f_{\rm{BH}}$ increases
approaching unity as $\beta \rightarrow 1$. These fractions are very
different from those of flat space, which can be calculated
analytically by integrating the appropriate photon angular
distribution [$\propto \gamma^{-2}(1-\beta \mu)^{-2}$] over the
cosine $\mu$ of the polar angle $\theta$ from $\mu = -1$ to $\mu =
0$ (for $f_{\infty}$), from $\mu =0 $ to $\mu_{\rm c} =
h/\sqrt{1+h^2}$ (for $f_{\rm{D}}$)  and from $\mu_{\rm c}$ to $\mu
=1$ (for $f_{\rm{BH}}$) where $\mu_{\rm c}$ is the cosine of the
angle that the horizon subtends at the distance of the source. In
the flat geometry one expects the maximum disk illumination to be
obtained for $\theta \sim 3 \cos^{-1}(\mu_c)$ or $\beta \simeq
\cos[3 \cos^{-1}(\mu_c)] \simeq 0.85$, a value substantially
different from that given in Figure~\ref{fig:fraction2}, the
difference attributed to the focusing effects of the geometry. Also,
the ratio of $f_{\rm{D}}$ to $f_{\infty}$, a quantity relevant for
the EW of a line, achieves in flat space its maximum value for
$\beta \simeq 0.65$.
%
%

\subsection{The Iron Line Profiles}

\begin{figure}[t]
\epsscale{1} \plottwo{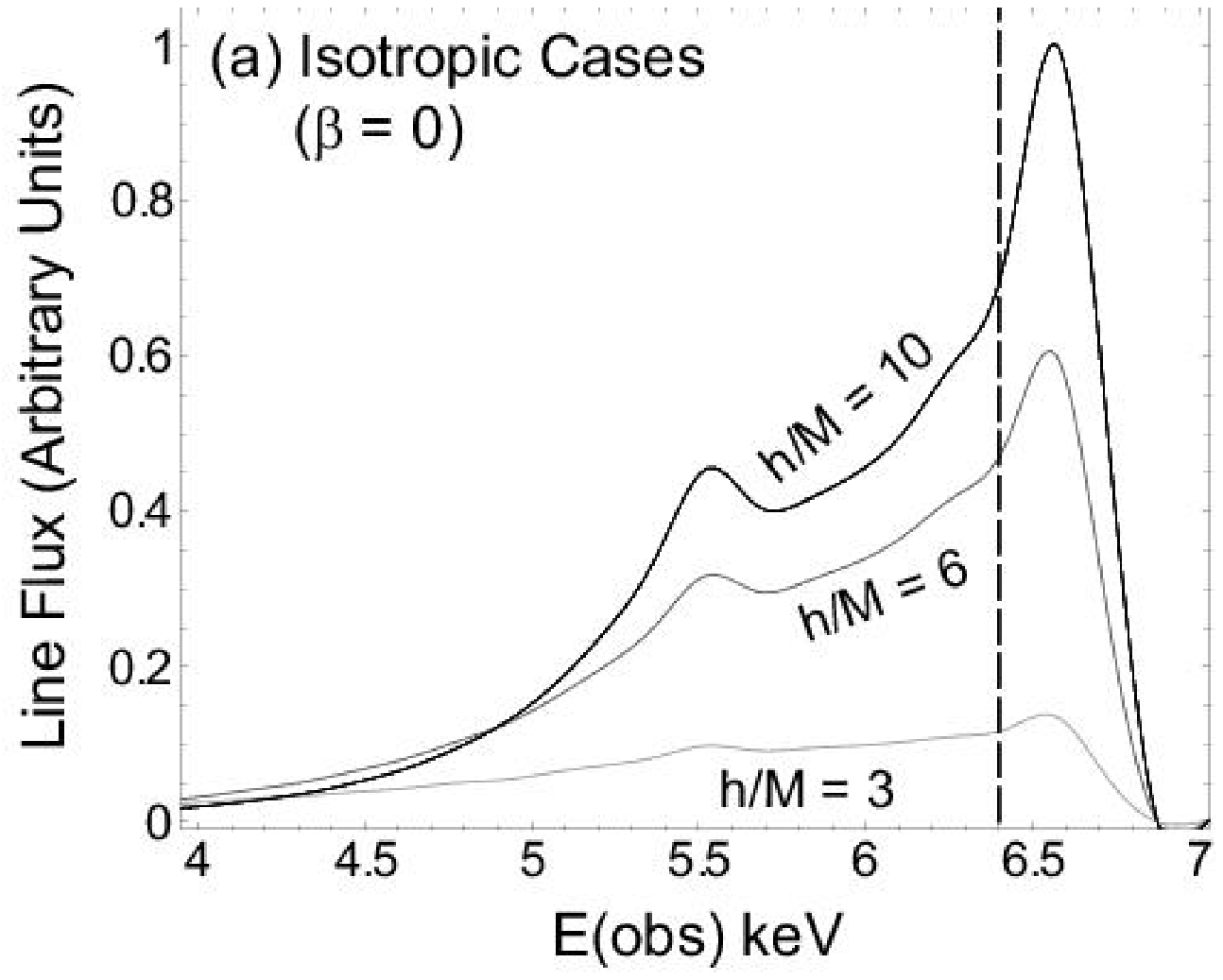}{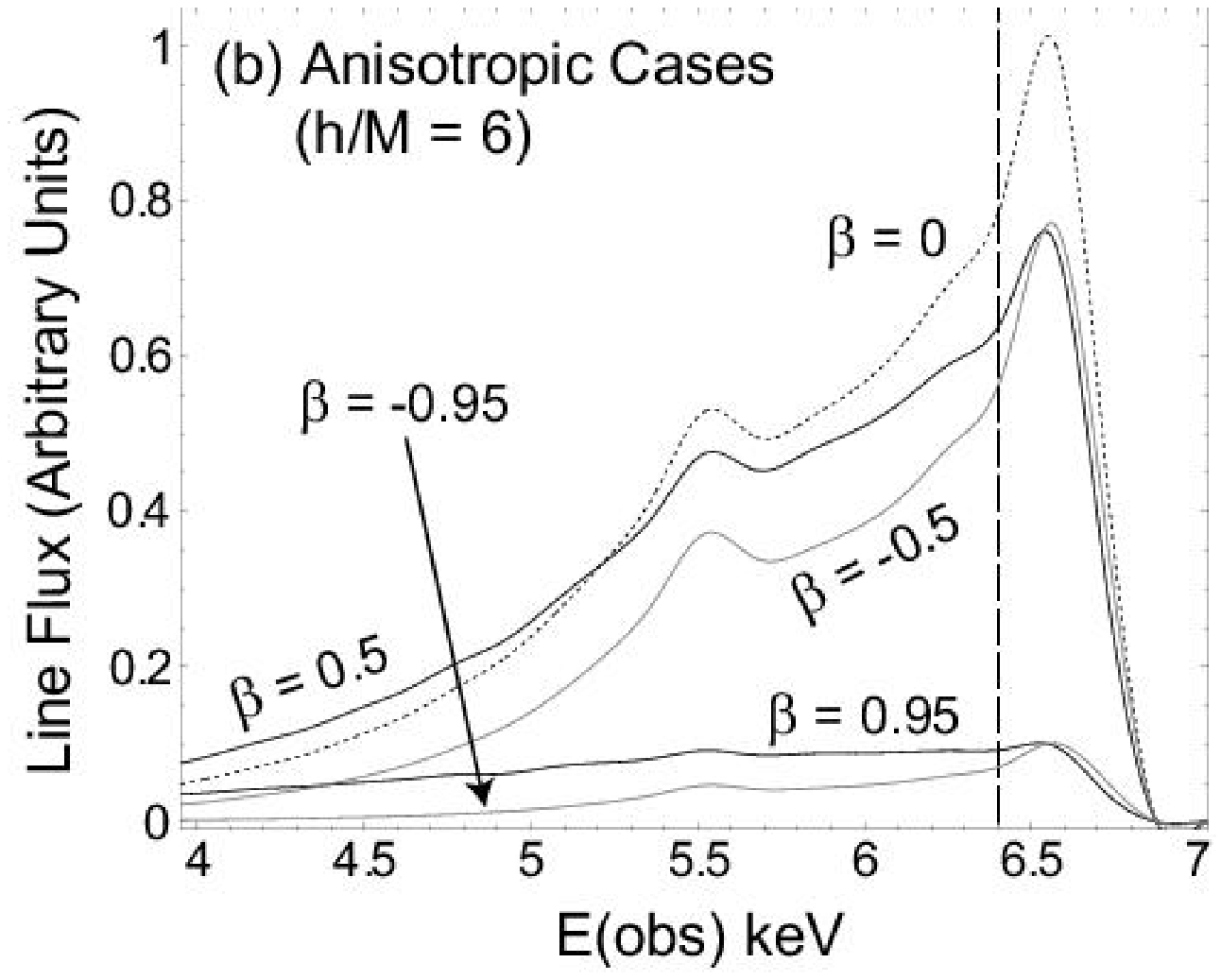} \caption{Iron line spectra
(in arbitrary units) for (a) isotropic and (b) anisotropic sources
with $a/M=0.99$.
    In (a) we show $h/M=3$, $6$ and $10$ for $\beta=0$ while in (b)
    we use $\beta=-0.95, -0.5, 0, 0.5$ and $0.95$ for $h/M=6$.
    Vertical dashed line denotes a cold iron line at 6.4 keV in the rest frame. }
\label{fig:line}
\end{figure}

In this section we produce the relativistic Fe line profiles
corresponding to the illumination laws derived above. The procedure
followed for that is the standard ray-tracing method given in the
literature \citep[e.g.,][]{Fanton97,Cadez98,Beckwith04,FT04}, by
which the photon orbits are followed from the source to the
observer's field of view (or image plane). In all cases the
inclination angle of the observer is 30$^{\circ}$, relevant for
Seyfert 1 galaxies, and the energy resolution is assumed to be
$\Delta E =150$ eV. {We only consider regions outside the ISCO for
producing line photons.} In Figure~\ref{fig:line} sample line
spectra are shown for (a) isotropic and (b) anisotropic sources. In
(a) the line profiles are shown for $h/M=3, ~6$ and $10$ to
illustrate the effects of changing the source height on the line
profile for $\beta=0$. In (b) we show the line profiles for a source
at a fixed height $h/M=6$ but of varying anisotropy given by
$\beta=-0.95,\; -0.5, ~0, ~0.5$ and $0.95$ to exhibit the effects of
source anisotropy on the corresponding line profiles.

In (a) the profiles are generally broader the closer the source is
to the disk due to the increased focusing of the photons toward the
disk inner edge. Because the line profiles were produced by
collecting photons only out to disk radius $r_{\rm out}/M = 30$, the
line flux decreases with decreasing $h/M$; this is because for a
decrease in $h/M$, the photon flux at $r/M \gsim 10$, which
contributes to the narrow peak of the line near 6.6 keV, decreases,
while it increases for the part of the disk that contributes to the
low ($E < 5 $ keV) emission. At the same time, however, as shown in
Figure~\ref{fig:fraction1} the fraction of the photons escaping to
infinity also decreases and as a result the line EW will not
necessarily decrease (provided that the point source is the only
source of illumination).

In (b) the calculated line spectra become broader with increasing
anisotropy of an approaching source ($0 < \beta < 1$), simply
because more photons are concentrated near the inner edge of the
disk. For sufficiently high values of $\beta>0$, the anisotropy
beams a large fraction of the photons towards the black hole
horizon, resulting in a decrease of the line absolute flux. The red
wing of the line, on the other hand, is not suppressed with non-zero
flux at energies less than 4 keV.
 However, as discussed above, the number of
photons escaping to infinity (the observer) also decreases and so
the EW of the line may not necessarily be negligible.
Figure~\ref{fig:line}b exhibits also, as a reference, the line
profile in the same geometry produced by sources receding from the
disk ($\beta <0$). The beaming effect in this case occurs in the
opposite direction (away from the black hole) suppressing the
overall line flux at all energies.



\begin{figure}[t]
\epsscale{1} \plottwo{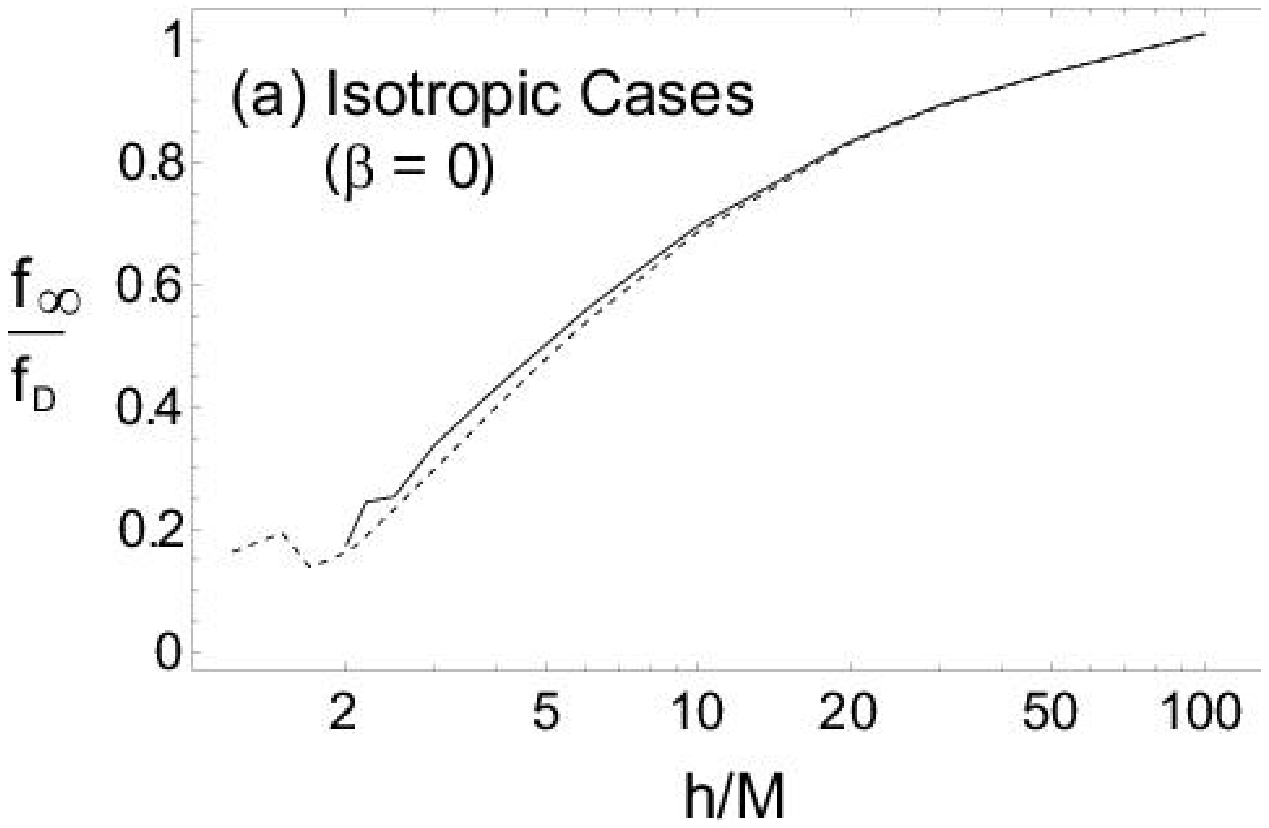}{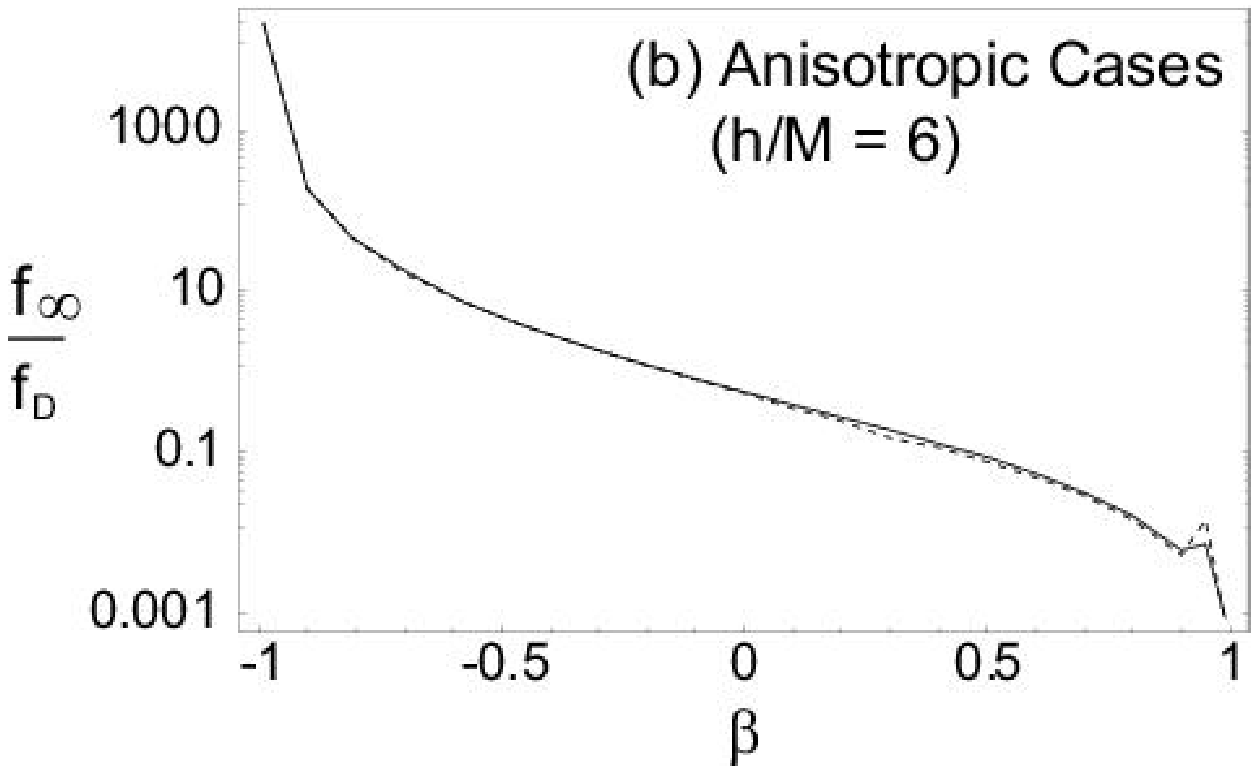} \caption{The ratio of the
fractions of the source photons escaping to infinity to that
    intercepted by the disk $f_\infty/f_{\rm{D}}$ for (a) an isotropic source (given as a function
    of the source height $h/M$) and (b) for an anisotropic source (given as a function of the
    anisotropy velocity $\beta$). Panel (a) corresponds to the flux ratio of Figure~\ref{fig:fraction1}
    while (b) corresponds to the flux ratio of
    Figure~\ref{fig:fraction2}. Solid lines are for $a/M=0$ while
    dotted ones for $a/M=0.99$. }
\label{fig:EW}
\end{figure} 

To provide an estimate of the influence of the source position and
anisotropy on the line EW we present in Figure~\ref{fig:EW} the
ratio of the photons escaping to infinity to those intercepting the
disk $f_\infty / f_{\rm{D}}$ as a function of (a) the source height
$h/M$ for the case of an isotropic source and (b) the anisotropy
parameter $\beta$ for the case of an anisotropic source at $h/M=6$,
i.e. cases that correspond to Figures~\ref{fig:fraction1} and
\ref{fig:fraction2}, respectively. While both fluxes are reduced as
the source approaches the black hole (or its anisotropy increases)
the ratio of the fluxes intercepted by the disk and escaping to
infinity actually increases. All other things remaining equal (e.g.
efficiency of Fe line production), this should lead to an increase
in the line EW.

\section{Discussion \& Conclusions}

We have produced through direct integration of a large number of
photon orbits the illumination law of thin accretion disks around
black holes by point--like sources at different heights above the
black hole on the axis of symmetry. We have also produced analytic
fitting formulae that can provide accurate approximations to the
illumination laws (in number flux) under the above
assumptions for both \sw and Kerr black holes. The corresponding
energy flux illumination profiles can be obtained by multiplying those
by the enhancement factor, ${\cal{D}}^{1+\alpha}$ approriate for each
situation considered. These can be employed as needed to
calculate the corresponding line profiles in a fashion similar to
that shown in \S 3.3. We have also produced for each case the
corresponding fractions of photons emitted by the source that escape
to infinity, intercept the disk or being absorbed by the black hole,
quantities necessary in producing the relative normalization between
the escaping (observed) flux and that reprocessed into either the Fe
line or the Compton reflection feature. Our overall results can be
summarized as follows:

(a) For sufficiently large values of $h/M~ (\gsim 50$) the
illumination profiles of isotropic sources approach those of flat
space, with the focusing effects of the geometry becoming noticeable
for $r/M \lsim 7$. As the height $h$ approaches the size of the
event horizon, the focusing effects of the geometry concentrates the
photon paths to the inner edge of the accretion disk while reducing
the illumination of larger radii; these features bear direct
relation to the shapes of the resulting line profiles.

(b) In addition to the focusing effects of the geometry, of interest
in the production of Fe line photons by the incident radiation is
also the increased rate of photon reception and the photon energy
shift between the emitting source and the receiving plasma, on a
Keplerian orbit on the accretion disk. Assuming that the disk and
line production is important only for radii larger than that of
ISCO, this amplification is significant for an extreme ($a/M=0.99$)
Kerr geometry but not for a \sw one. As shown in
Figures~\ref{fig:IL-1} and \ref{fig:IL-2} the main effect of the
hole spin parameter $a$ is to reduce the size of the horizon and the
size of ISCO, which simply determines the minimum radius at which
the illumination (and hence the contribution to the line profile)
needs to be calculated. This is a rather surprising outcome of our
calculations; despite the different topologies of the orbits with
increasing black hole spin $a$ the resulting illumination laws
appear to change very little for $h/M \gsim 3$ (however the profiles
become significantly different for smaller values of $h$). As a
result the fitting formulae we provide in \S 3.1 (with the
modifications for the rate and photon energy discussed in \S 3.1)
should suffice to produce illumination profiles by isotropic sources
for both \sw and Kerr black holes.

(c) Motivated by the need of illumination more concentrated to the
disk inner radius, in order to explain certain particularly broad Fe
lines, we have modeled the illumination of disks by point-like
anisotropic sources. We modeled the source anisotropy by
Lorentz-boosting an isotropic source along the symmetry axis and
toward the black hole by a constant velocity $\beta$. We found that
for sufficiently large values of $\beta>0$, the illumination law (in
number flux) assumes the shape of a single power law $\propto
r^{-3}$, over the entire range of radii. We have also found that
further increase in $\beta$ does not modify this functional form but
simply decreases the overall normalization because of absorption of
most of the photons by the black hole. While the black hole spin
does not affect the illumination profile it does affect the ratio of
source photons intercepted by the disk (which can contribute to the
line emission) to those that escape to infinity and contribute to
the continuum emission, in particular in the case of anisotropic
source illumination.


(d) We found that the photon energy flux illumination law, obtained
by multiplication of the photon number flux by the factor
${\cal{D}}^{1+\alpha}$, can be significantly steeper in the
innermost disk region of a Kerr geometry with an index that approach
the value $q \sim 4$ for a small range in radius.

Our calculations indicate that it is hard to obtain disk
illumination profiles more concentrated than those discussed above
(demanded by certain observations) within the confines of our
assumptions. However, modification of these assumptions should
easily allow for much steeper profiles. A point-like source on the
symmetry axis is an idealization that simplifies the computational
task. A point-like source off axis will greatly enhance the
illumination in the region immediately underneath its position; this
requires a more complicated non-axisymmetric description, not
demanded as yet by the observations (but it may in the future). It
is also apparent that an extended source can produce arbitrarily
steep illumination profiles depending on the assumption of the
radial distribution of the X-ray intensity within the source. In the
absence of any physically compelling prescription, it is simplest to
approximate the illumination as a combination of power-laws: (a)
outer illumination $\propto r^{-3}$ appropriate for large distances;
our results indicate this to be a valid assumption for distances
much larger than the vertical extent of the source and (b) a steeper
one appropriate for the illumination of the inner portions of the
disk. While the computation of the resulting illumination is
possible under an extended source assumption by extending the
procedure described in this note we do not believe that such a model
would lead to any significant insights about the physics of these
sources.

To this point we have not discussed at all the nature of the
illuminating source. Several proposals have been made in the
literature of which we provide an incomplete list: An accretion disk
corona \citep[e.g.,][]{Haardt91}, powered either by magnetic fields
threading the disk or by magnetic fields threading the black hole
\citep[e.g.,][]{Li02,Reynolds04}. An alternative is the emission by
shocks in the accretion flows that funnels (through magnetic fields)
the disk material onto the black hole \citep[e.g.,][for
magnetohydrodynamic shocks in accretion]{Fukumura07}. The source can
be either stationary or variable, as such shocks were proposed to be
unstable and oscillating \citep[see, e.g.,][for discussion of shock
stability]{Trussoni88,Aoki04}. Finally, we have not addressed at all
the issue of disk ionization, an issue that affects its line
producing efficiency \citep[e.g.,][]{Nayakshin00,Ballantyne01} as
they are not related directly to the main thrust of the present
work.


Assuming that the plunging region (inside the ISCO) is
optically-thin due to very low mass-accretion rate, photons emitted
from the source could in principle cross the plunging region. We
have followed photon trajectories (from the source to the equatorial
plane) and found that certain of them turned around (due to strong
light bending) to hit the bottom side of the disk at larger radii
and contribute to the illumination there; we found this contribution
to the illumination to be unimportant.

We have produced in \S 3 fitting formulae for the illumination of
accretion disks (in photon number flux), which can be easily
converted into the photon energy flux needed in the calculations of
Fe line profiles. Our models and fits are consistent with the values
implied by spectral fits to the Fe lines of most AGNs. Assuming
power-law distributions $\propto r^{-q}$ for emissivity, the values
of the index $q$ obtained for MCG--5-23-13, NGC 4051 and NGC 3516 in
recent {\it Suzaku} observations \citep[][]{Reeves06} are $q \simeq
2-3$; our models can accommodate easily the lower values of $q$
while demanding a reasonably high degree of anisotropy for values $q
\gtrsim 3$. However, the disk illumination under the assumptions
considered in this note cannot account for the width of the Fe line
profile of MCG --6-30-15 as obtained either by {\it XMM-Newton}
\citep{Wilms01,Fab02} or more recently by {\it Suzaku}
\citep{Miniutti07} which demand a very steep ($q \simeq 4.6$)
illumination of the inner disk section along with a flatter one ($q
\simeq 2.6$) at larger radii. As we discussed earlier, the easiest
scheme that could accommodate such a profile would be an extended
($r/M \simeq 2$) rather than the point-like source used in our
models. An example of such a source has been given in
\citet{Miniutti03}, however we believe that other types of extended
sources could provide good fits as well. Another possibility has
also been proposed that magnetically-induced torque (on the disk)
could produce very steep emissivity profiles \citep[e.g.,][]{Li02}.
Modeling such sources goes beyond the scope of the present work and
maybe undertaken in the future.

\acknowledgments

We would like to thank the anonymous referee for a number of useful
and insightful suggestions. K.F. was supported in part by an
appointment to the NASA Postdoctoral Program at the Goddard Space
Flight Center, administered by Oak Ridge Associated Universities
through a contract with NASA.


\section{Appendix A}

In the Appendix A we present the orbit equations which we have
integrated in producing Figures~\ref{fig:path} and
\ref{fig:beaming1} in \S 3. The symmetry of the problem defines an
axis which in general is different from the axis connecting the
source to the black hole. In this case, in addition to the total
angular momentum ${\cal L}$ of the photon with respect to the black
hole, there is also its component about the hole's rotation axis
$L_z$. Therefore, in this case the photon has two degrees of
freedom, i.e. the impact parameters with respect to the black hole
$\eta \equiv {\cal L}/E^2$ and the impact parameter with respect to
the hole's rotation axis $\xi \equiv L_z/E$.

In this case the equations of motion in the $r$ and $\theta$
directions  are
\begin{equation}
\dot r^2 = \frac{R(r)}{\Sigma^2} ~~~~ {\rm and} ~~~~ \dot \theta^2 =
\frac{\Theta(\theta)}{\Sigma^2} \ ,
\end{equation}
where $\Sigma \equiv r^2 + a^2 {\rm cos}^2 \theta$ and $\Delta \equiv
r^2 - 2Mr + a^2$, with $a$ being the angular momentum per unit mass
associated with the Kerr metric, and the functions $R(r)$ and
$\Theta(\theta)$ are defined by
\begin{eqnarray}
R(r) &\equiv & [(r^2+a^2)E - a L_z]^2 - \Delta[{\cal L} + (L_z -
aE)^2] \ \label{Rr} , \\
\Theta(\theta) & \equiv & {\cal L} - (-a^2 E^2 + L_z^2 \, {\rm
cosec}^2 \theta) \, {\rm cos}^2 \theta  \ . \label{RTh}
\end{eqnarray}
In analogy with equation (\ref{integral}) these are supplemented by
the following equations for $\dot t$ and $\dot \phi$
\begin{equation}
\dot t \equiv \frac{dt}{d \tau} = 2 \frac{(r^2+a^2)[(r^2+a^2)E - a
L_z] + a \Delta L_z - a^2 E \Delta \,{\rm sin}^2 \theta}{\Delta (2
r^2 + a^2 + a^2 {\rm cos}2 \theta)} \ , \\
\end{equation}

\begin{equation}
\dot \phi  \equiv  \frac{d \phi}{d \tau}  = 2 \frac{a[a^2 E - a L_z
+ E(r^2 - \Delta)] + L_z \Delta \, {\rm cosec}^2 \theta}{\Delta (2
r^2 + a^2 + a^2 {\rm cos}2 \theta)} \ .
\end{equation}
%
By dividing equations (\ref{Rr}) and (\ref{RTh}) above by $E^2$ and
re-expressing them in terms of the impact parameters $\eta$ and
$\xi$ we obtain
\begin{eqnarray}
R(r) & = & [(r^2+a^2) - a \xi]^2 - \Delta[\eta + (\xi -
a)^2] \ , \\
\Theta(\theta) & = & \eta - (-a^2 + \xi^2 {\rm cosec}^2 \theta)\,
{\rm cos}^2 \theta \ .
 \label{RTh2}
\end{eqnarray}

Specializing to a source located on the axis of symmetry, $\xi =0$;
then the equations of motion and the integrals of motion read
respectively as
\begin{eqnarray}
\Sigma^2 \dot r^2      & = &(r^2+a^2)^2 - \Delta(\eta + a^2) \ , \label{Krdot}
\\
\Sigma^2 \dot \theta^2 & = &\eta + a^2 \,{\rm cos}^2 \theta \ ,
\label{Kthetdot}
\end{eqnarray}
and
\begin{eqnarray}
\dot t & =& 2 \frac{(r^2+a^2)^2 - a^2 \Delta \,{\rm sin}^2
\theta}{\Delta (2r^2 + a^2 + a^2 \, {\rm cos} \, 2 \theta)} \ ,
\\
\dot \phi & = & 2 \frac{a ( a^2 + r^2 - \Delta)}{\Delta (2r^2 + a^2
+ a^2 \, {\rm cos} \, 2 \theta)} \ .
\end{eqnarray}
%
The angle between the axis perpendicular to the disk and the
direction of photon emission (we call it $\delta$ instead of $\psi$
in this case) is given by an expression equivalent to that of
equation (\ref{angle})
\begin{equation}
{\rm cot} \delta = \frac{\vert g_{rr} \vert ^{1/2} \dot r}{\vert
g_{\theta \theta} \vert ^{1/2} \dot \theta} = \frac
{1}{\Delta^{1/2}} \left[\frac{(r^2+a^2)^2 - \Delta (\eta +
a^2)}{\eta + a^2 \, {\rm cos}^2 \theta} \right]^{1/2} \ ,
\end{equation}
which can be readily inverted to provide an expression for the
impact parameter $\eta$ in terms of $\delta$, namely
\begin{equation}
\eta(\delta, \theta; r, a) = \frac{1}{\Delta} \left[(r^2 + a^2)^2 -
a^2 \Delta (1 + {\rm cos}^2 \theta \, {\rm cot}^2 \delta ) \right]
{\rm sin}^2 \delta \ , \label{Kerrangle2}
\end{equation}
an equation that reduces to equation (\ref{angle2}) for $a=0$.

We follow a procedure similar to that of the \sw case, i.e. we opt
for the numerical integration of second-order equations for $r$ and
$\theta$. Being emitted from a point along the symmetry axis
($\xi=0$) the photons still have one degree of freedom namely the
impact parameter $\eta$. We use equation (\ref{Kerrangle2}) to
relate the local emission angle $\delta$ to $\eta$ and
equations~(\ref{Krdot}) and (\ref{Kthetdot}) to obtain the initial
values of $\dot r$ and $\dot \theta$ necessary for the integrations.
The integrations proceed again by a choice of the angle $\delta$
chosen uniformly between $0$ and $\pi$ from $(r=h, ~\theta = 0)$
until $\theta = \pi/2$ and $r$ is greater than the black hole
horizon corresponding to the specific value of the hole angular
momentum $a$. For completeness we present below the second-order
equations used in the integrations. These are the following:
\begin{equation}
\ddot r = \frac{1}{\Sigma^2} \left[ (a^2 {\rm sin} \,2 \theta \,
\dot \theta - 2 \, r \dot r) \dot{r}\Sigma + 2 (r^2 + a^2) \, r - (r
- M ) (\eta + a^2) \right] \ ,
\end{equation}
and
\begin{equation}
\ddot \theta = \frac{1}{2\Sigma^2} \left[ 2(a^2 {\rm sin} \,2 \theta
\, \dot \theta - 2 \, r \dot r)\dot \theta \Sigma  - a^2 {\rm sin}
\, 2 \theta  \right] \ .
\end{equation}
The general case for a photon source off the axis, $\xi \neq 0$, is straightforward
to obtain using the general form of the equations for $R(r)$ and $\Theta(\theta)$.
However, in this case one needs to choose an additional angle in the (local to
the source) azimuthal direction which has to be related to $\xi$ in a similar
fashion. In that case the illumination pattern depends also on the azimuthal angle
$\phi$ and the results sufficiently complicated to be presented within this work.

\section{Appendix B}

As emitted photons hit the accretion disk, its local energy and the
rate at which they are intercepted by the disk will be
relativistically shifted. In order to calculate photon energy flux
measured in fluid/disk comoving frame, one must take those effects
into account. Photon redshift factor from an emitter's frame to a
receiver's frame is defined as
\begin{equation}
{\cal{D}} \equiv \frac{(\mathbf{u} \cdot \mathbf{p})_{\rm
r}}{(\mathbf{u} \cdot \mathbf{p})_{\rm e}} = \frac{u^{\mu}_{\rm r}
p_{\mu}}{u^{\mu}_{\rm e} p_{\mu}} \ , \label{eq:redshift-1}
\end{equation}
where ``r" and ``e" respectively denote the receiver and emitter.
For photon's four-momentum, we have $\mathbf{p} = (-E,\pm E
\sqrt{R(r)}/ \Delta, \pm \sqrt{\Theta(\theta)}, \xi E)$ where $E$ is
the photon energy in the emitters frame
\citep[][p.347]{Chandrasekhar83}. For four-velocity we generally
assume $\mathbf{u}=(u^t, u^r, 0 , u^\phi)$. Thus, we obtain
\begin{equation}
{\cal{D}} = \left(\frac{u^t_{\rm r}}{u^t_{\rm e}}\right)
\frac{(1-\xi \Omega \mp \sqrt{R(r)} v^r / \Delta)_{\rm r}}{(1-\xi
\Omega \mp \sqrt{R(r)} v^r / \Delta)_{\rm e}} \ ,
\label{eq:redshift-2}
\end{equation}
where $v^r \equiv u^r/u^t$ and $\Omega \equiv u^\phi / u^t$. For
example, for photons emitted from a stationary on-axis source
($\xi=0$) reaching the accretion disk (outside ISCO), we find
\begin{equation}
{\cal{D}} = \frac{u^t_{\rm disk}}{u^t_{\rm source}} =
\left[\frac{(g_{tt})_{\rm source}} {(g_{tt} + 2 g_{t\phi} \Omega_K +
 g_{\phi \phi} \Omega_K^2)_{\rm disk}}\right]^{1/2} \ , \label{eq:redshift-3}
\end{equation}
where $\Omega_K \equiv \sqrt{M}/(a \sqrt{M} + r^{3/2})$ is Keplerian
frequency of the disk \citep[][p.336]{Chandrasekhar83}.

\clearpage

\end{document}